\documentclass[longauth]{aa}  

\usepackage{graphicx}
\usepackage{multirow}
\usepackage{amssymb}
\usepackage{makecell}
\usepackage{tabularx}
\usepackage{float}
\usepackage{placeins}
\usepackage{stfloats}
\usepackage{tipa}

\newcommand\setItemnumber[1]{\setcounter{enumi}{\numexpr#1-1\relax}}

\usepackage{txfonts}
\usepackage{natbib}
\bibpunct{(}{)}{;}{a}{}{,} 
\usepackage{hyperref}
\hypersetup{
    colorlinks=true,
    citecolor=blue,
    filecolor=magenta
}
\begin{document}

   \title{The COSMOS-Web deep galaxy group catalog up to $z=3.7$}

   \author{Greta Toni
          \inst{1,2,3}\thanks{\email{greta.toni4@unibo.it}}
          \and
          Ghassem Gozaliasl\inst{4,5}
          \and
          Matteo Maturi\inst{3,6}
          \and
          Lauro Moscardini\inst{1,2,7}
          \and
          Alexis Finoguenov\inst{5}
          \and
          Gianluca Castignani\inst{2}
          \and
          Fabrizio Gentile\inst{1,2,8}
          \and
          Kaija Virolainen\inst{4}
          \and
          Caitlin M. Casey\inst{9,10}
          \and
          Jeyhan S. Kartaltepe\inst{11}
          \and
          Hollis B. Akins\inst{9}
          \and
          Natalie Allen\inst{10,12}
          \and
Rafael C. Arango-Toro\inst{13}
\and
Arif Babul\inst{14,15}
\and
Malte Brinch\inst{16}
\and
Nicole E. Drakos\inst{18}
\and
Andreas L. Faisst\inst{19}
\and
Maximilien Franco\inst{8}
\and
Richard E. Griffiths\inst{18,20}
\and
Santosh Harish\inst{11}
\and
Günther Hasinger\inst{21,22,23}
\and
Olivier Ilbert\inst{13}
\and
Shuowen Jin\inst{10,17}
\and
Ali Ahmad Khostovan\inst{11,24}
\and
Anton M. Koekemoer\inst{25}
\and
Maarit Korpi-Lagg\inst{4}
\and
Rebecca L. Larson\inst{11}
\and
Jitrapon Lertprasertpong\inst{11}
\and
Daizhong Liu\inst{26}
\and
Georgios Magdis\inst{10,12,17}
\and
Richard Massey\inst{27}
\and
Henry Joy McCracken\inst{28}
\and
Jed McKinney\inst{9}
\and
Louise Paquereau\inst{28}
\and
Jason Rhodes\inst{29}
\and
Brant E. Robertson\inst{30}
\and
Mark Sargent\inst{31}
\and
Marko Shuntov\inst{10,12}
\and
Masayuki Tanaka\inst{32}
\and
Sina Taamoli\inst{33}
\and
Elmo Tempel\inst{34}
\and
Sune Toft\inst{10,12}
\and
Eleni Vardoulaki\inst{35}
\and
Lilan Yang\inst{11}
}

   \institute{Dipartimento di Fisica e Astronomia "A. Righi", Alma Mater Studiorum Universit\`a di Bologna, via Gobetti 93/2, 40129 Bologna, Italy; \email{greta.toni4@unibo.it}
\and
INAF - Osservatorio di Astrofisica e Scienza dello Spazio di Bologna, via Gobetti 93/3, 40129 Bologna, Italy
\and
Zentrum für Astronomie, Universit\"at Heidelberg, Philosophenweg 12, 69120 Heidelberg, Germany; \email{maturi@uni-heidelberg.de}
             \and
             Department of Computer Science, Aalto University, PO Box 15400, 00076, Espoo, Finland; \email{ghassem.gozaliasl@aalto.fi}
             \and
             Department of Physics, University of Helsinki,
              Gustaf H\"allstr\"omin katu 2, 00560 Helsinki, Finland
              \and
             ITP, Universität Heidelberg, Philosophenweg 16, 69120 Heidelberg, Germany
              \and
             INFN - Sezione di Bologna, Viale Berti Pichat 6/2, 40127 Bologna, Italy; \email{lauro.moscardini@unibo.it}
             \and
             CEA, IRFU, DAp, AIM, Université Paris-Saclay, Université Paris Cité, Sorbonne Paris Cité, CNRS, 91191 Gif-sur-Yvette, France
             \and The University of Texas at Austin, 2515 Speedway Blvd Stop C1400, Austin, TX 78712, USA \and Cosmic Dawn Center (DAWN), Denmark
             \and Laboratory for Multiwavelength Astrophysics, School of Physics and Astronomy, Rochester Institute of Technology, 84 Lomb Memorial Drive, Rochester, NY 14623, USA
             \and 
             Niels Bohr Institute, University of Copenhagen, Jagtvej 128, DK-2200, Copenhagen, Denmark
             \and
             Aix Marseille Univ, CNRS, CNES, LAM, Marseille, France 
\and
Department of Physics and Astronomy, University of Victoria, 3800 Finnerty Road, Victoria, BC  V8P 5C2
Canada
\and
Infosys Visiting Chair Professor, Department of Physics,  Indian Institute of Science, Bangalore, 560012 India
\and
Instituto de Física y Astronomía, Universidad de Valparaíso, Avda. Gran Bretana˜ 1111, Valparaíso, Chile
\and DTU-Space, National Space Institute, Technical University of Denmark, Elektrovej 327, DK-2800 Kgs. Lyngby, Denmark
\and 
 Department of Physics \& Astronomy, University of Hawaii at Hilo, 200 W. Kawili Street, Hilo, HI 96720, USA
 \and
Caltech/IPAC, MS 314-6, 1200 E. California Blvd. Pasadena, CA 91125, USA
\and
Department of Physics, Carnegie Mellon University, 5000 Forbes Avenue, Pittsburgh, PA 15213, USA
\and
TU Dresden, Institute of Nuclear and Particle Physics, 01062 Dresden, Germany 
\and
DESY, Notkestrasse 85, 22607 Hamburg, Germany
\and 
Deutsches Zentrum für Astrophysik, Postplatz 1, 02826 Görlitz, Germany
\and
 Astrophysics Division, NASA Goddard Space Flight Center, Greenbelt, MD 20771, USA
\and
Space Telescope Science Institute, 3700 San Martin Dr., Baltimore, MD 21218, USA
\and
Purple Mountain Observatory, Chinese Academy of Sciences, 10 Yuanhua Road, Nanjing 210023, China
\and
Department of Physics, Centre for Extragalactic Astronomy, Durham University, South Road, Durham DH1 3LE, UK
\and 
Institut d’Astrophysique de Paris, UMR 7095, CNRS, and Sorbonne Université, 98 bis boulevard Arago, F-75014 Paris, France
\and
Jet Propulsion Laboratory, California Institute of Technology, 4800 Oak Grove Drive, Pasadena, CA 91001, USA
\and
Department of Astronomy and Astrophysics, University of California, Santa Cruz, 1156 High Street, Santa Cruz, CA 95064, USA
\and
International Space Science Institute (ISSI), Hallerstr. 6, CH-3012 Bern, Switzerland
\and
National Astronomical Observatory of Japan, 2-21-1 Osawa, Mitaka, Tokyo 181-8588, Japan
 \and
Department of Physics and Astronomy, University of California, Riverside, 900 University Ave, Riverside, CA 92521, USA
\and
Tartu Observatory, University of Tartu, Observatooriumi 1, 61602 T\~oravere, Estonia
\and
Th\"{u}ringer Landessternwarte, Sternwarte 5, 07778 Tautenburg, Germany
}

   \date{Received -; accepted -}

  \abstract
   {Galaxy groups with total masses below $\sim 10^{14} \, M_\odot$ and up to a few tens of members are the most common galaxy environment, marking the transition between the field and the most massive galaxy clusters. In this framework, identifying and studying groups plays a crucial role in understanding structure formation and galaxy evolution. Despite the challenges in detecting such relatively small structures, modern deep surveys allow us to build well-characterized samples of galaxy groups up to the regime where the structures we observe today were taking shape.}
   {We aim to build the largest deep catalog of galaxy groups to date over the COSMOS-Web field effective area of 0.45 deg$^2$. }
   {We leveraged the deep imaging, high resolution, and high-quality photometry from the James Webb Space Telescope observations of the COSMOS-Web field. We used the recent COSMOS-Web photometric catalog with sky position, photometric redshift, and magnitude in a reference band for each selected galaxy. We performed the group search with the Adaptive Matched Identifier of Clustered Objects (AMICO) algorithm, a linear matched filter based on an analytical model for the cluster/group signal. This algorithm has already been tested in wide and deep field surveys, including a successful application to COSMOS data up to $z=2$. In this work, we tested the algorithm's performances at even higher redshift and searched for protocluster cores and groups at $z>2$. To benchmark this relatively unexplored regime, we compiled a list of known protoclusters in COSMOS at $2 \leq z \leq 3.7$ and matched them with our detections. We studied the spatial connection between detected cores through a clustering analysis. We estimated the purity and the completeness of our group sample by creating data-driven mocks via a Monte Carlo approach with the SinFoniA code and linked signal-to-noise to purity levels to define desired purity thresholds.} 
   {We detected 1678 groups in the COSMOS-Web field up to $z=3.7$ with a purity level of $\sim 77\%$, providing a deep catalog of galaxy members that extends nearly two magnitudes deeper than the previous application of AMICO to COSMOS. Around 670 groups have been detected with a purity of 90\%. Our catalog includes more than 850 groups whose photometric redshift was confirmed by assigning robust spectroscopic counterparts.}
   {This catalog of galaxy groups is the largest ultra-deep group sample built on JWST observations so far and offers a unique opportunity to explore several aspects of galaxy evolution in different environments spanning $\sim$12 Gyr and study groups themselves, from the least rich population of groups to the formation of the most massive clusters.}


   \maketitle

\section{Introduction}
Galaxies are distributed in a complex structure known as the cosmic web, where the densest regions are connected by filaments and walls \citep[e.g.,][]{bond_how_1996}. Groups of galaxies, which reside in this large-scale environment, host the bulk of galaxies and play a crucial role in the understanding of their evolution, but also of the formation and evolution of structures like massive galaxy clusters themselves \citep[e.g.,][]{tully_nearby_1987,eke_galaxy_2004,lovisari_scaling_2021} and their cosmological impact \citep[e.g., see the review by][]{allen_cosmological_2011}. Galaxy groups represent the transition range in halo masses from massive galaxies to galaxy clusters and show different physical properties and evolutions with respect to galaxy clusters \citep[e.g.,][]{giodini_stellar_2009,mcgee_accretion_2009,mccarthy_case_2010,gaspari_agn_2011}. Despite there being no universally defined threshold to separate groups and clusters, they are generally distinguished in the literature by using limits, for instance, in mass at around $  \sim 10^{14} \, M_\odot$ or in the number of galaxy members at around 50 galaxies \citep[e.g.,][]{paul_understanding_2017, lovisari_scaling_2021}. Galaxies hosted in groups are believed to evolve in a different way with respect to galaxies in the field, undergoing substantial alteration, for instance, in star formation rate (SFR) \citep[e.g.,][]{scoville_evolution_2013,darvish_cosmic_2014,darvish_effects_2016,taamoli_cosmos2020_2024}, morphology \citep[e.g.,][]{mandelbaum_density_2006,capak_effects_2007,bamford_galaxy_2009}, and gas and metal content \citep[e.g.,][]{catinella_galex_2013,chartab_mosdef_2021}. This occurs via interactions \citep[][]{hausman_galactic_1978}, ram-pressure stripping \citep[][]{gunn_infall_1972} and a variety of other mechanisms \citep[see e.g.,][for a review]{boselli_environmental_2006}. Although some environmental effects are observed to be more efficient in dense massive clusters, some phenomena are strongly affecting galaxies in lower-density environments such as groups \citep[e.g.,][]{bianconi_locuss_2018,lietzen_environments_2012,vulcani_gasp_2018}, and even filaments \citep[e.g.,][]{laigle_cosmos2015_2018}.

Environmental effects in galaxy groups have been widely explored at $z \lesssim 1.5 $ \citep[e.g.,][]{wilman_galaxy_2005,george_galaxies_2012,salerno_filaments_2019,balogh_gogreen_2021,mcnab_gogreen_2021,reeves_gogreen_2021,baxter_when_2023,kukstas_gogreen_2023,gozaliasl_cosmos_2024}. However, the understanding of galaxy groups and galaxy evolution in groups becomes more complicated and unexplored at $z \gtrsim 1.5$ where the classical morphology-density relation observed in the local universe \citep{dressler_galaxy_1980} seems to fade and group and cluster galaxies exhibit properties more consistent with those of the field galaxies, in terms, for example, of SFRs and morphologies \citep[e.g.,][]{brodwin_era_2013,alberts_star_2016}. In addition to this, the interval $z = 1.5-2.0$ marks the transition between virialized objects observed locally ($z<1.5$) and the maturing phase of protoclusters ($2.0 <z<3.5$), the progenitors of galaxy clusters \citep[e.g.,][]{shimakawa_mahalo_2018}. The study of high-redshift galaxy groups and protostructures is key to understanding the evolution and star formation history of galaxies, their connection to the dark matter distribution, and the impact of physical processes such as AGN feedback \citep[e.g.,][]{tanaka_x-ray_2012,zhang_high-redshift_2020,kiyota_cluster_quiescent_2025}.

In the COSMOS field, several group catalogs have been produced over the years, like the X-ray-selected samples by \cite{finoguenov_xmm-newton_2007} and \cite{george_galaxies_2011}, which used the XMM-Newton data \citep{hasinger_xmm-newton_2007} to robustly identify galaxy groups up to $z\sim 1.0$. In a more recent effort, \cite{gozaliasl_chandra_2019} revised the X-ray catalogs by incorporating all available X-ray observations from Chandra and XMM-Newton in the 0.5-2 keV band and using photometric and spectroscopic data for the optical counterparts \citep[e.g.,][]{ilbert_cosmos_2009,mccracken_ultravista_2012,laigle_cosmos2015_2016}. 
Several other works have leveraged the availability of spectroscopic redshifts \citep{lilly_zcosmos_2007,lilly_zcosmos_2009} to detect groups up to $z \sim 1$ \citep[e.g.,][]{knobel_zcosmos_2012} and proto-groups up to $z \sim 3$ \citep[e.g.,][]{diener_proto-groups_2013}.

In this work, we made use of the deepest contiguous 0.54 deg$^2$ galaxy catalog available to create the largest deep galaxy group catalog created to date, extending from $z=0$ up to $z=3.7$. We used the COSMOS-Web photometric catalog of galaxies (Shuntov et al. in prep.), which is the result of the largest contiguous imaging of the sky performed with the James Webb Space Telescope (JWST). The COSMOS-Web survey \citep{casey_cosmos-web_2023} is a unique combination of the unprecedented depth and spatial resolution of JWST and the coverage and data availability of the COSMOS field \citep{scoville_cosmic_2007}, making it a key resource for the study and definition of the large-scale structures in the cosmic web over around 50 million Mpc$^3$ \citep{casey_cosmos-web_2022}.
The JWST NIRCam near-infrared (NIR) photometric coverage, combined with more than 30 photometric bands available for the COSMOS field (from ultraviolet to infrared) allows for high-quality photometric redshift (photo-$z$, hereinafter) estimation, with precision below $\sim $ 0.03 even at the faintest magnitudes in the redshift range we are interested in \citep{arango-toro_history_2024,shuntov_cosmos-web_2024}. This enables the creation of robust samples up to the protocluster regime ($z \gtrsim 2$) and a detailed study of structures potentially up to $z=7$ and beyond \citep[e.g.,][]{morishita_protocluster_jwst_2023, morishita_earlyoverdensity_2024}.

In order to create such a deep galaxy group catalog we made use of the Adaptive Matched Identifier of Clustered Objects \citep[AMICO;][]{bellagamba_amico_2018,maturi_amico_2019}, an algorithm based on a linear optimal matched filter \citep{maturi_optimal_2005, bellagamba_optimal_2011} which extracts signal maximizing the signal-to-noise ratio ($S/N$) and without any explicit color selection of galaxies. Without requiring spectroscopic information or galaxy colors, AMICO is able to detect clusters and groups up to high redshift and down to low masses \citep[e.g., up to $z = 2$ and down to less than $10^{13}$ M$_\odot$ as it was shown by][]{toni_amico-cosmos_2024}. In its simplest application, AMICO detection is based on the spatial and luminosity distribution of galaxies in clusters without using color information, which limits the possibility of biases related to the presence (or absence) of the cluster red sequence, particularly important when moving to high redshifts. The algorithm is one of the two cluster finders selected for the ESA's Euclid mission \citep{laureijs_euclid_2011,mellier_unveiling_2018}, given its performances in terms of completeness and purity when tested on Euclid-like mock catalogs \citep{euclid_collaboration_euclid_2019}. AMICO has already been validated and successfully applied to several surveys such as the Kilo Degree Survey \citep[KiDS;][]{maturi_amico_2019} and the miniJPAS \citep{maturi_minijpas_2023}.

More recently, \citet{toni_amico-cosmos_2024} utilized the AMICO algorithm to produce a cluster and group catalog containing 1269 candidates in total in the range $0.1 <z <2.0$. This catalog was the result of three independent AMICO runs performed using the magnitude in three different photometric bands as reference galaxy property, and to compute the luminosity function of clusters. 490 candidates were consistently detected in all three runs. A comparative analysis of the three runs suggested a possible cut in signal-to-noise to define a more robust sample. X-ray properties were assigned to these detections by cross-matching with the group sample by \cite{gozaliasl_chandra_2019}. For unmatched detections, the X-ray properties were estimated using the same Chandra and XMM-Newton data. The final catalog includes 622 candidate clusters and groups with optical properties, X-ray flux estimates, and estimates of mass. This large sample of candidates with assigned X-ray properties enabled the calibration of scaling relations between two AMICO mass proxies (i.e., richness and amplitude) and X-ray mass, permitting the study of their redshift dependence and the selection of the most stable photometric bands.

\begin{figure*}
    \centering
    \includegraphics[width=18cm]{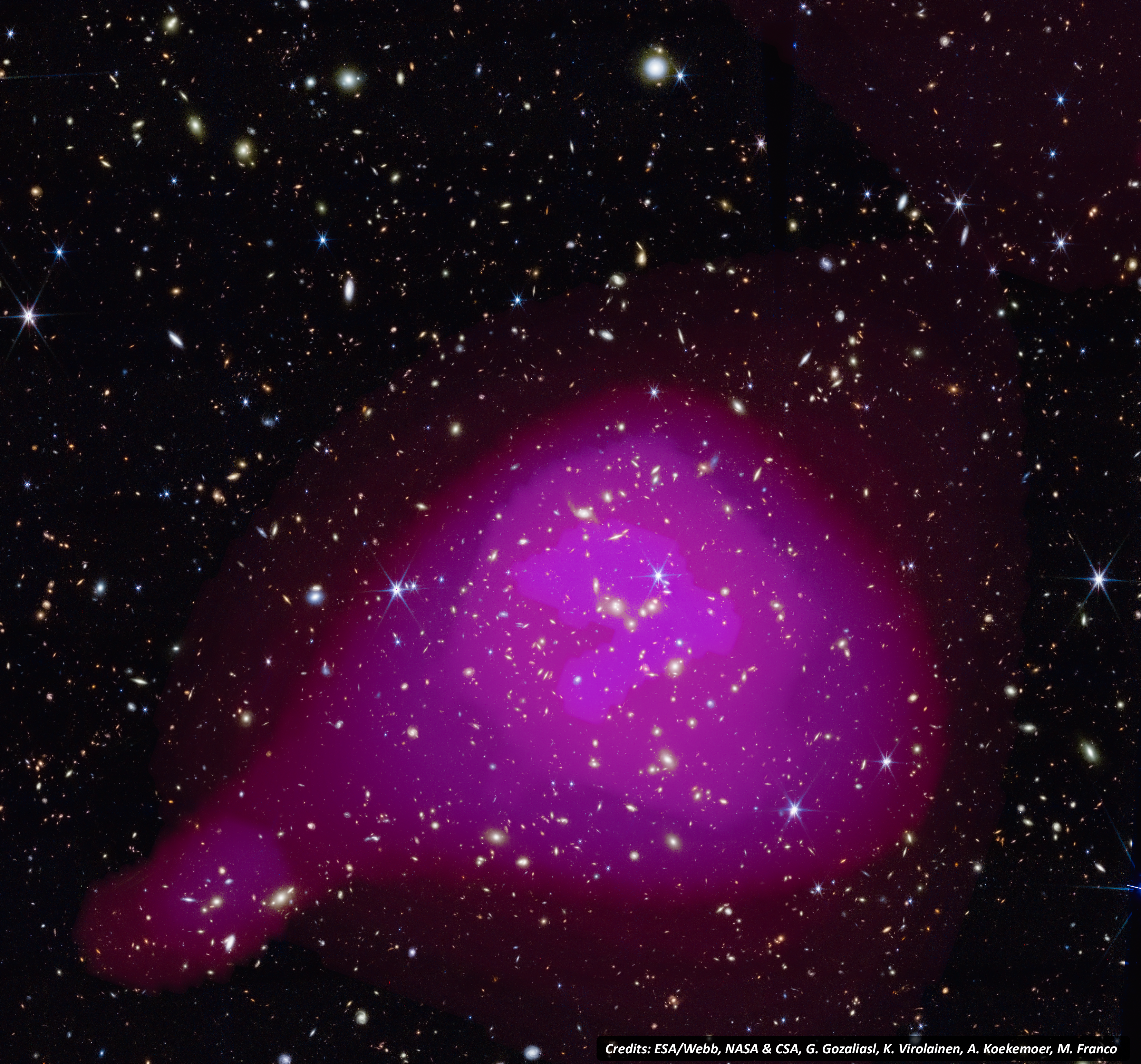}
    \caption{JWST $rgb$ (F444W as $r$, [F150W, F277W] as $g$, F115W as $b$) color-composite image of the most massive group in the COSMOS-Web field. The JWST image is overlapped with the X-ray extended emission (pink) from the combined XMM-Newton and Chandra 0.5–2 keV wavelet-filtered image.}
    \label{fig:overlap_xray}
\end{figure*}

In the work we present in this paper, we leveraged the insights and experience gained from the successful AMICO-COSMOS group search described in \cite{toni_amico-cosmos_2024} combined with the high-accuracy, deep photometric redshifts provided by the COSMOS-Web survey to present the largest deep galaxy group sample detected to date, produced by applying the AMICO algorithm to the COSMOS-Web data. Figure \ref{fig:overlap_xray} shows the impressive resolution of the JWST color-composite image overlaid with the X-ray extended emission \citep{gozaliasl_chandra_2019}, for one of the richest and most massive groups in the COSMOS field.

This paper is structured as follows.
Section \ref{dataset} introduces the COSMOS-Web data, the galaxy catalog, and the selection criteria applied to create a robust input dataset for the group search. Section \ref{amico} outlines the main steps of the detection process and the core principles of the AMICO algorithm. In Sect. \ref{results}, we present the results of the group search, first by introducing the group catalog and then by discussing the creation of realistic mocks and the evaluation of the purity of the sample. In Sect. \ref{highz}, we examine detections at $z>2$, compare them to known objects, and analyze their clustering properties. Finally, Sect. \ref{conclusions} summarizes our key findings, their implications, and the potential future research based on this work.

Throughout this paper, we use the term ``group'' to refer to our candidate detections, given that only a few objects in the analyzed field are expected to have masses larger than $10^{14}$ M$_\odot$ or more than 50 members according to previous detections performed in the COSMOS field \citep[e.g.,][]{knobel_zcosmos_2012,gozaliasl_chandra_2019,toni_amico-cosmos_2024}. We assume a standard concordance flat $\Lambda$CDM cosmology with $\Omega_\mathrm{m}=0.3$, $\Omega_\Lambda =0.7$, and $h=H_0$ / (100 km/s/Mpc) $=0.7$.

\section{Data}\label{dataset}
\subsection{Photometric catalog}

The COSMOS-Web Survey \citep[PIs: J. Kartaltepe and C. Casey;][]{casey_cosmos-web_2023}, is a 255-hour Cycle 1 observation program conducted using JWST. The survey spans 0.54 deg$^2$ and utilizes four NIRCam filters \citep{rieke_performance_2023}, achieving a 5$\sigma$ point-source depth between 27.5 and 28.2 mag. 
Observations were carried out using the F115W, F150W, F277W, and F444W filters \citep{casey_cosmos-web_2023} and for a non-contiguous 0.19 deg$^2$ area imaged in the F770W MIRI filter \citep{wright_james_2022}, with a 5$\sigma$ point-source depth between 25.3 and 26.0 mag.

The COSMOS field \citep{scoville_cosmic_2007, capak_first_2007} benefits from a rich legacy of multi-wavelength data, from X-rays to radio \citep[e.g.,][]{civano_chandra_2016, hasinger_xmm-newton_2007, smolcic_vla-cosmos_2017}. Optical data include the $u$-band observations from CFHT's MegaCam \citep{sawicki_cfht_2019}, the high-resolution data from the Hubble Space Telescope (ACS-HST) in the F814W band \citep{koekemoer_cosmos_2007}, the Hyper-Suprime-Cam (HSC) imaging in the $g$, $r$, $i$, $z$, and $y$ bands, in addition to the 13 intermediate and narrow bands of Subaru Suprime-Cam (SC) \citep{taniguchi_subaru_2015}. The UltraVISTA survey \citep{mccracken_ultravista_2012, moneti_vizier_2023} completes the coverage at NIR-wavelengths with the $Y$, $J$, $H$, and $Ks$ bands, which are complementary to JWST's NIRCam and MIRI.

The new COSMOS-Web photometric catalog has been developed, targeting specifically the portion of the COSMOS field observed by JWST, succeeding previous COSMOS galaxy catalogs \citep[e.g.,][]{ilbert_mass_2013, laigle_cosmos2015_2016, weaver_cosmos2020_2022}. More than 784,000 sources have been detected over the 0.54 deg$^2$ area, using a PSF-homogenized $\chi^2$ detection image, which combines all NIRCam filters. Source extraction is challenging due to the variation in PSF sizes across space- and ground-based facilities, ranging from 0.05 to 1.0 arcseconds. To address this, \texttt{SourceXtractor++} \citep{bertin_sourcextractor_2022} has been used to model Sérsic surface brightness profiles \citep{sersic_influence_1963} at NIRCam resolution, followed by photometric extraction in each band. The resolution of JWST images enables the separation of previously blended sources \citep{arango-toro_history_2024}. A more detailed description of the COSMOS-Web photometric catalog used in this study is provided by Shuntov et al. in preparation and \cite{arango-toro_history_2024}.

\subsection{Photometric redshifts}

Accurate photometric redshift estimation is crucial for reliably detecting galaxy groups, as the precision of these redshifts directly affects the identification and characterization of such structures. Photometric redshifts in the COSMOS-Web source catalog are computed with the template-fitting code \texttt{LePhare} \citep{arnouts_measuring_2002, ilbert_accurate_2006}, with an expanded parameter space for the template library allowed by the depth and coverage offered by this field. The template library consists of 12 templates based on stellar population synthesis models \citep{bruzual_stellar_2003}, with 42 different ages and including various star formation histories (SFHs) and metallicities ($Z = 0.008Z_\odot$, $0.02Z_\odot$) as outlined by \citet{ilbert_evolution_2015}.

In order to assess photometric redshift accuracy, the COSMOS-Web redshifts have been compared to more than 11,000 spectroscopic redshifts (spec-$z$s, hereinafter), with confidence level  $>$97\%, from the compilation created by \cite{khostovan_spec_compil_2025} \citep[e.g.,][]{lilly_zcosmos_2007, lilly_zcosmos_2009, kartaltepe_multiwavelength_2010, kartaltepe_rest-frame_2015, silverman_fmos-cosmos_2015, kashino_fmos-cosmos_2019}. The photometric redshift precision is around 0.01 for $m_{F444W} < 24$, with a 2\% rate of catastrophic failures\footnote{Photometric redshift performance is estimated in terms of the $\sigma_{NMAD}$ indicator, that is $\sigma_{NMAD} = 1.48 \times median[(|\Delta z - median(\Delta z)|)/(1 +  z_{spec})]$, where $\Delta z = z_{phot} - z_{spec}$, the difference between photometric and spectroscopic redshift. Galaxies are defined as outliers if $|\Delta z | > 0.15(1+z_{spec})$.}. Even for the faintest galaxies at $26 < m_{F444W} < 28$ and for high redshifts in the interval we are interested in ($z < 4$), the precision remains better than 0.03 with $\sim $10\% failure \citep{arango-toro_history_2024,shuntov_cosmos-web_2024}. 

\subsection{Data cleaning and visibility mask }\label{mask}

\begin{figure}
   \centering
   \includegraphics[width=9cm]{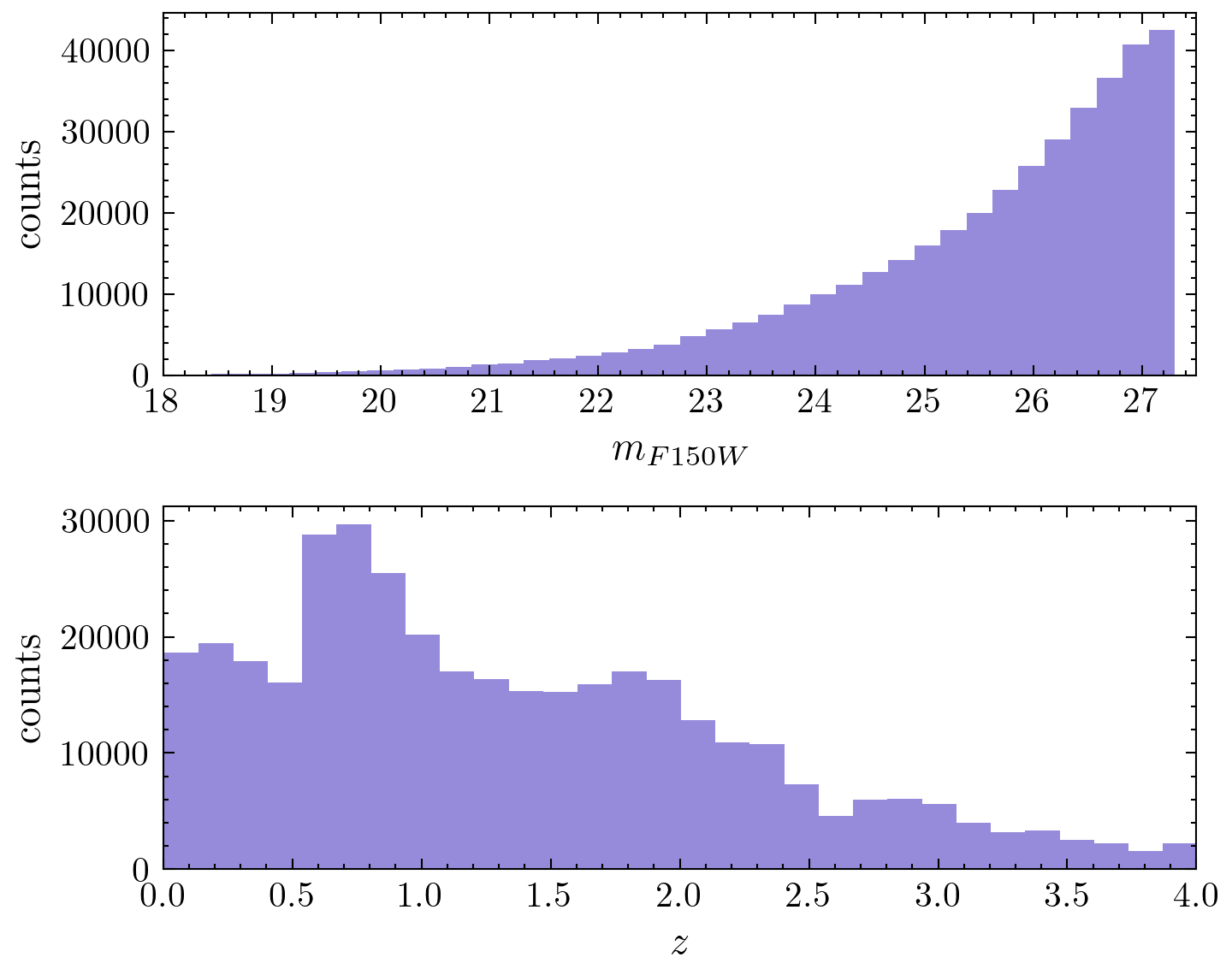}
      \caption{Distribution in magnitude (top panel) and redshift up to  $z=4$ (bottom panel) of the cleaned galaxy catalog used as input for the group search.}
         \label{distri}
\end{figure}

We cleaned the data set by keeping only galaxies (including active galaxies) based on \texttt{LePhare} classification. We removed masked objects limiting the selection to both $\texttt{FLAG\_STAR\_JWST=0}$ and $\texttt{FLAG\_STAR\_HSC=0}$ to ensure not only the availability and reliability of JWST photometry but also of the external ground-based photometry which is used to estimate photometric redshifts. Removing unsafe regions for ground-based data implies a loss of area of around $10\%$ compared to the area selected only considering JWST photometry quality. We decided to sacrifice this area in exchange for the improvement in photo-$z$ uncertainties and in physical properties resulting from SED fitting that including good-quality ground-based photometry ensures \citep[see][and Shuntov et al. in preparation, for further details]{shuntov_cosmos-web_2024}. A cleaned and high-quality sample is crucial for the study of the clustering of galaxies in three dimensions, while on the contrary, the inclusion of badly characterized galaxies with uncertain and inaccurate photo-$z$s can contaminate the sample with spurious detections, for instance, in correspondence with artifacts. For this reason, we additionally kept only the galaxies with the best photo-$z$ quality flag, $\texttt{LP\_warn\_fl=0}$, to reject artifacts like snowballs and sources with inconsistent photometry between different bands. We then cut the catalog at the mode of the magnitude distribution in the F150W band, which we chose as the reference band (see Sect. \ref{amico}). This minimizes the introduction of noise and spurious detections in the catalog and defines the depth of the galaxy catalog, that is $m_{F150W}=27.3$.
As a reference, this catalog is almost 2 magnitudes deeper than the one we used in \cite{toni_amico-cosmos_2024} and more than 3 magnitudes deeper than the cluster search performed in KiDS (\citeauthor{maturi_amico_2019} \citeyear{maturi_amico_2019}; Maturi et al. in prep.). To further reduce the contamination of the galaxy sample due to poor classification, we removed all objects with an extremely small radius, choosing as a threshold radius $\sim 0.01^{\prime\prime}$, which is the value that divides into two distinct groups the sources in the $m_{F150W}-$radius plane. The total number of selected galaxies used for the group detection is 389248. The redshift and magnitude distribution of the cleaned input galaxy catalog is shown in Fig. \ref{distri}.

In order to include in our group detection the information about the footprint of the survey and the areas of the field that are inaccessible or potentially contaminated during the observation phase, we need to generate a so-called visibility mask. Using a visibility mask as input in the detection process has a twofold purpose: it filters out unsafe sources excluding them from the input catalog and it allows the detection algorithm to account for inaccessible areas during the detection procedure. We based the visibility mask on the input galaxy catalog, using an approach similar to that used in \cite{toni_amico-cosmos_2024}. In particular, we created the visibility mask starting by assigning the value 1 (masked) to all mask pixels devoid of galaxies in the selected catalog described above. This ensures areas lying outside the field and areas affected by bright star halos are accounted for. Then, we built polygons to cover areas that may be affected by star spikes. To do this, we used the Incremental Data Release of the HSC bright-star masks by \citet{coupon_bright-star_2018}, extracted from Gaia DR2 \citep{gaia_collaboration_gaia_2018}, with magnitude $G<18$. The spike and halo size follow an exponential relation with the magnitude of the star, with the same approach presented by \cite{coupon_bright-star_2018}, and the polygons were converted to binary masks using the \texttt{venice} code \citep{coupon_venice_2018}. Besides these polygons, we visually inspected the galaxy density maps in different redshift bins and manually masked areas with possible artifacts close to the star halos and/or field borders. During the visual inspection, due to suspiciously shaped overdensities at $z>2$, we also discarded the area occupied by the central galaxies of a $z\sim 0.1$ known group in COSMOS, which is listed, for instance, by \cite{gozaliasl_chandra_2019} as ID20149. This makes this known bright object undetectable but improves purity at higher redshifts. This masking procedure ensures our galaxy catalog is as clean as possible from spurious or bad-photometry detections, especially for high-$z$ detections. 

The resulting composite visibility mask yielded an effective area in which the group search was performed of about 0.45 deg$^2$.

\section{Detection of groups with AMICO}\label{amico}
AMICO \citep{bellagamba_amico_2018, maturi_amico_2019} detects galaxy clusters and groups in photometric galaxy catalogs using position, photometric redshift, and any additional galaxy property. The algorithm is based on a linear optimal matched filter \citep[e.g.,][]{maturi_optimal_2005} that extracts a signal for which we have an a priori model. 
The galaxy density can be described as the sum of a signal component and a noise component, representing the cluster and the field galaxies, respectively. The signal component $S$ is expressed by $S(\boldsymbol{x}) = A M_c(\boldsymbol{x})$ where $M_c(\boldsymbol{x})$, is the a priori cluster model and a function of the $n$ galaxy properties contained in the vector $\boldsymbol{x}$. The factor $A$, the signal amplitude, is retrieved as a convolution of the data with an optimal filter defined via constrained minimization and is related to the cluster richness. AMICO generates a three-dimensional amplitude map and selects detections at the peaks with the highest signal-to-noise ratios. Then, it attributes membership probability to galaxies whenever a candidate is identified. This probability is then used to compute the apparent richness, $\lambda$, as the sum of all member probabilities, and the intrinsic richness, $\lambda_\star$, considering only members restricted to $m_\star+1.5$ and inside $R_{200}$, where $m_\star$, the luminosity function knee magnitude, and $R_{200}$, the virial radius, are model parameters, as we describe below and in \citet{toni_amico-cosmos_2024}. In this paper, we focus on the specific characteristics of this particular application of AMICO to the new COSMOS-Web data. Therefore, for a complete description of the mathematical formalism and working principle of AMICO, we refer the reader to \cite{bellagamba_optimal_2011,bellagamba_amico_2018} and \cite{maturi_amico_2019}.

For this specific application, we adopted the same amplitude-map resolution used for the AMICO-COSMOS catalog \citep{toni_amico-cosmos_2024}, that is $\Delta r =0.3^\prime$ (on the sky plane) and $\Delta z =0.01$.
We used AMICO with a single galaxy property, which we chose to be the magnitude in the JWST F150W band (simply $m$, hereinafter). This photometric band, in addition to having good resolution and low background noise with respect to other bands, was proven to be stable to the calibration of the mass-proxy scaling relation at high redshift observed in the $H$-band run described in \cite{toni_amico-cosmos_2024}. For what concerns the redshift probability distribution $p(z)$ of each galaxy, we decided to rely on an analytical Gaussian distribution that peaks at the redshift with the highest probability for each galaxy with its $1\sigma$ uncertainties.

\begin{figure}
   \centering
   \includegraphics[width=9cm]{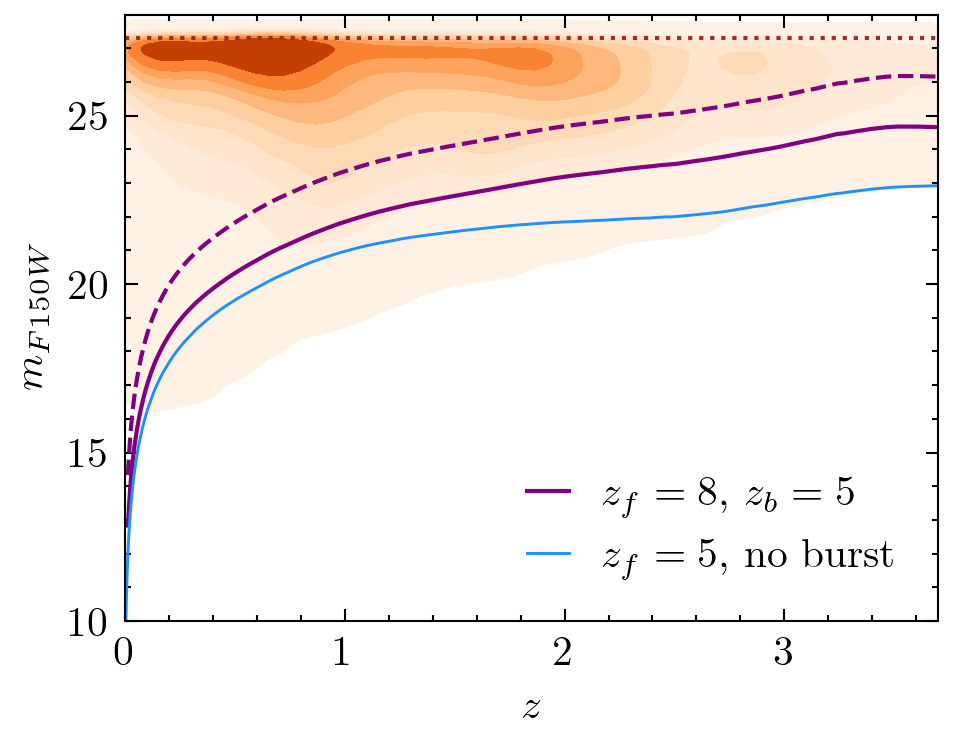}
      \caption{Magnitude and redshift of the galaxies of the input catalog (orange density contours) and evolution with redshift of the characteristic magnitude of the luminosity function, $m_\star$, for two different models with different formation redshifts and star formation burst, as shown in the legend. The model with $z_f=8$ and past burst is the one used for the model in this work, marked by the solid purple line. This model better describes the trend of magnitude with respect to the one represented by the solid blue line. The dashed purple line indicates the same model, but for $m_\star+1.5$ which is the limit used in the definition of the intrinsic richness. }
         \label{mstar_z}
\end{figure}

\noindent\textit{Cluster model.}
To describe the expected distribution of cluster galaxies, we built an analytical model with a truncated Navarro-Frenk-White profile \citep{navarro_universal_1997} and a Schechter luminosity function \citep{schechter_analytic_1976}. We used the following parameters from the literature: normalization as estimated by \cite{hennig_galaxy_2017}, concentration parameter from \cite{ragagnin_cosmology_2021} and faint-end slope of the luminosity function from \cite{andreon_jkcs_2014}. The characteristic magnitude of the luminosity function, $m_\star$, was estimated using evolutionary synthesis models with GALEV \citep{kotulla_galev_2009}. We used the same GALEV configuration as in \cite{toni_amico-cosmos_2024}, evolving a massive ($\sim 10^{11} \, M_\odot$) elliptical galaxy. Figure \ref{mstar_z} shows the redshift evolution of $m_\star$ overlapped with the selected galaxies (orange density contours) for a model with formation redshift, $z_f=8$ and an exponentially declining star formation burst (purple line), and one with $z_f=5$ and without burst (light blue line). The former marks the exponential cut-off in the number of galaxies and better describes the magnitude evolution trend of the data set up to at least $z \sim 3.7$, redshift to which we extended our search. This first $m_\star$ evolution trend was chosen to build the cluster model. We adopted a resolution in magnitude of $\Delta m = 0.5$ for the model and the noise describing group and field galaxies, respectively. This ensures sufficient galaxy statistics in each magnitude bin.

\noindent\textit{Noise model.} We estimated the noise, which describes the distribution of field galaxies, by approximating it to the general distribution of the whole galaxy sample. Given the small area covered by this group search, we investigated the impact of group galaxies on the noise and observed visible peaks localized in redshift, which could be due to physical overdensities. We cleaned the noise model from the group galaxy contamination by dividing the field into four non-overlapping quadrants of roughly the same area and by taking the median noise value of each corresponding noise bin for the four quadrants as the final one. This cancels out the contribution of localized overdensities that are present in a specific quadrant and not in the others. Additionally, we performed a regularization of the noise model similar to the one described in \cite{toni_amico-cosmos_2024}, attributing a large value to empty bins and those with $m<m_\star-3$.

\begin{figure*}
   \centering
   \includegraphics[width=18cm]{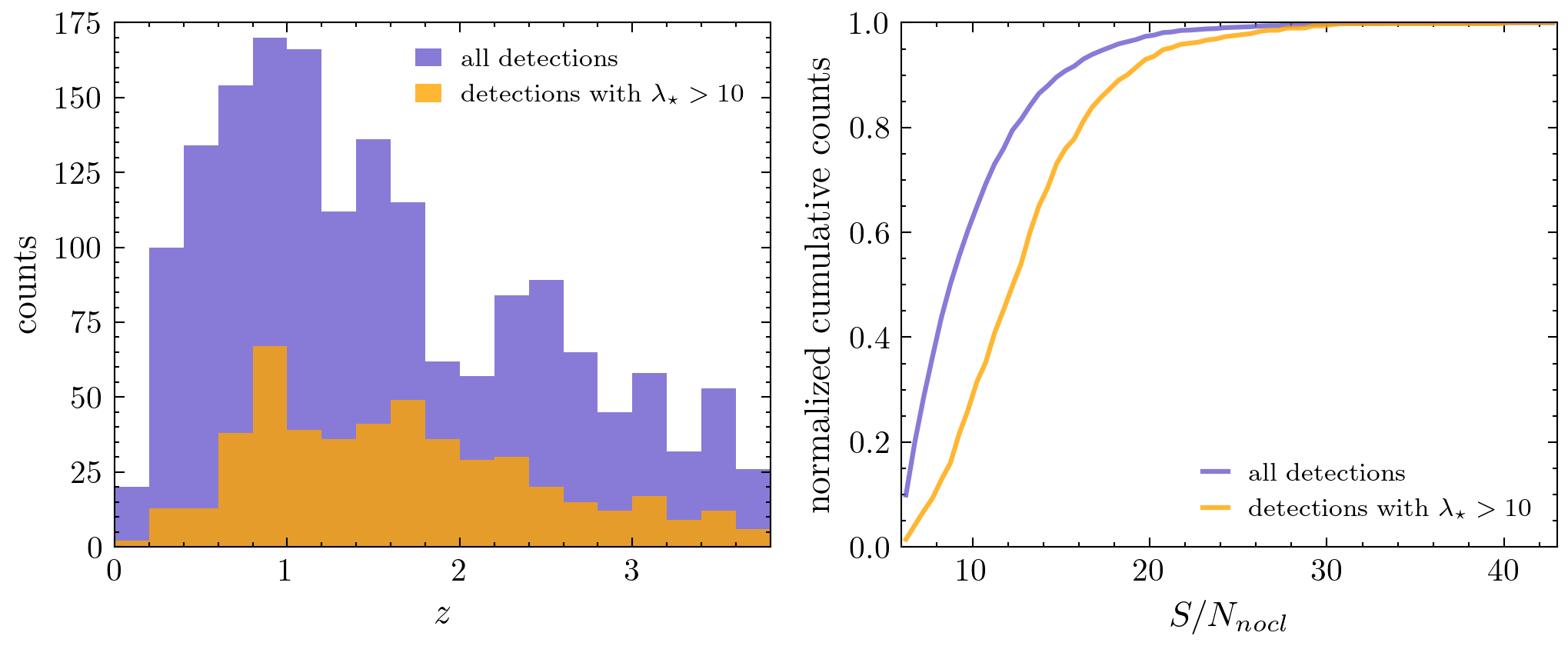}
      \caption{Redshift distribution (left panel) and normalized cumulative signal-to-noise distribution (right panel) for all the detections in the COSMOS-Web group catalog. In orange, we show the richest detections, with $\lambda_\star >10$.}
         \label{statprop}
\end{figure*}

\begin{figure}
   \centering
   \includegraphics[width=9cm]{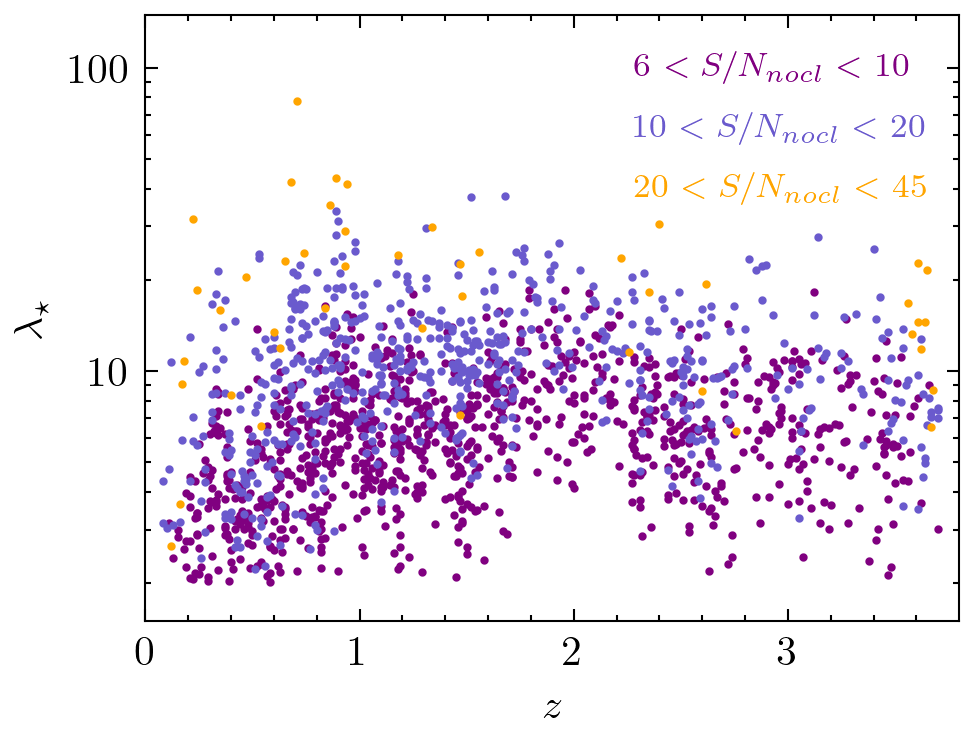}
      \caption{Intrinsic richness, $\lambda_\star$, for the sample of detected groups and its trend with redshift, in three different $S/N_{nocl}$ bins as indicated in the plot.}
         \label{ls}
\end{figure}
\begin{figure*}
    \centering
    \includegraphics[width=1\linewidth]{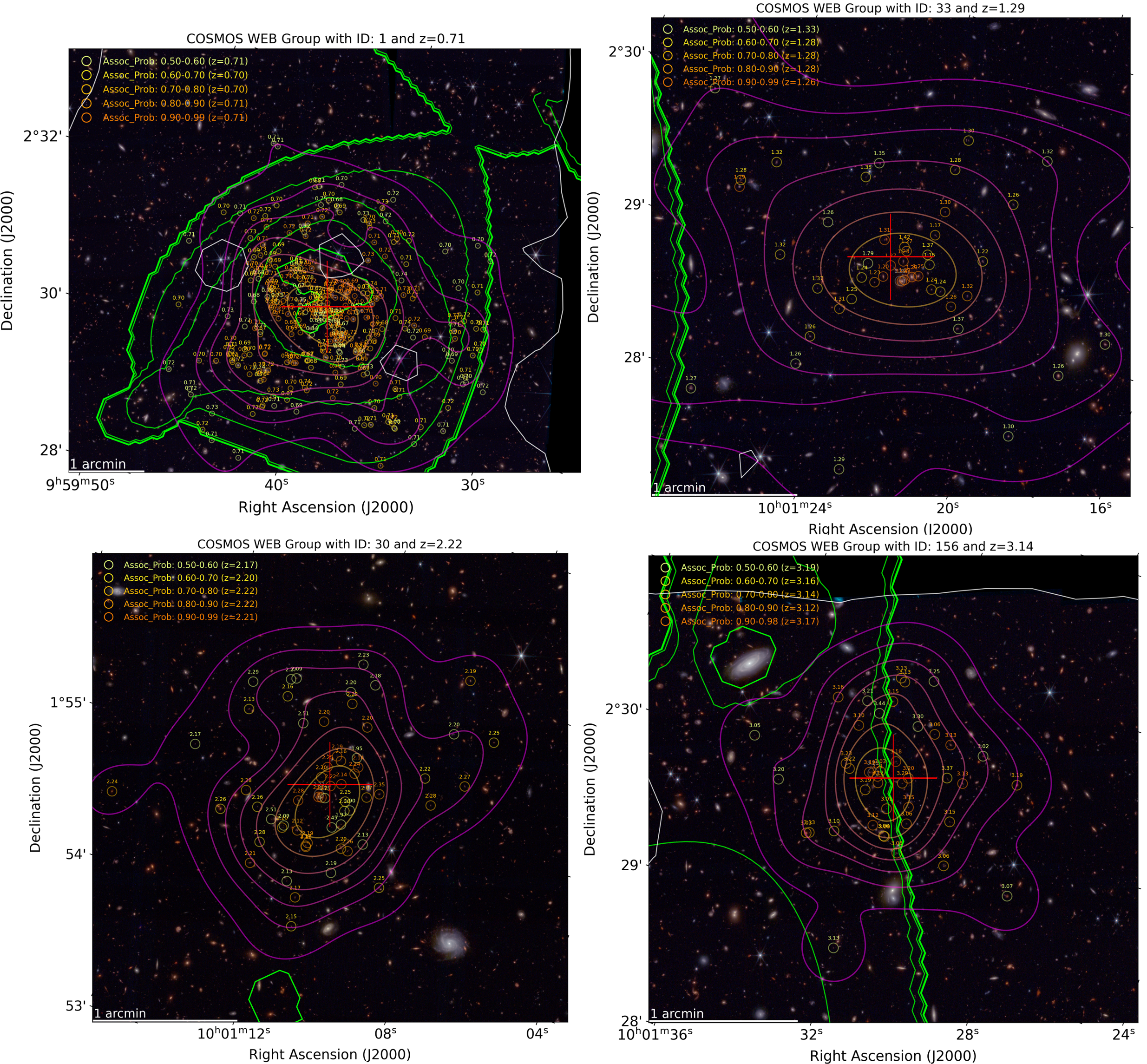}
    \caption{Four examples of detections present in the group catalog, at different redshifts. Circles indicate member galaxies, color-coded with membership probability and with their redshift printed next to each circle. Purple-to-yellow contours mark the density of galaxies and green contours the X-ray emission from the combined XMM-Newton and Chandra mosaic image in the 0.5--2 keV band. White lines delimit masked areas. The center of the group is marked with a red cross.  In the top left panel, a group at $z=0.71$ with $\lambda_\star \sim 78$ and central AGN (the same group shown in Fig. \ref{fig:overlap_xray}). In the top right panel, a group at $z=1.29$ with $\lambda_\star \sim 14$. In the bottom left panel, a group at $z=2.22$ with $\lambda_\star \sim 24$ and on the bottom right panel one at $z=3.14$ with $\lambda_\star \sim 28$, flagged for being on the extended X-ray emission of a low-$z$ group and for being at the edge of the field, as visible on the top of the stamp. JWST $rgb$ images are in the same filters as in Fig. \ref{fig:overlap_xray}.}
    \label{fig:panelexamplpe}
\end{figure*}
\section{Catalog of galaxy groups}\label{results}

We performed a group search with AMICO over the effective area of 0.45 deg$^2$ in the COSMOS-Web field, as previously described. We ran the code down to $(S/N_{nocl})_{min}=6.0$, where $S/N_{nocl}$ is the AMICO signal-to-noise ratio, which does not include shot-noise from cluster/group members in the amplitude variance. When taken as a reference, this was proved to yield a more stable redshift dependence of the purity of the sample, with respect to the standard $S/N$, that includes both background and member contribution (Maturi et al. in prep.). For further details on the definition and on the difference between the two signal-to-noise ratios in AMICO, we refer the reader to \cite{bellagamba_amico_2018} and \cite{maturi_amico_2019}. We then cut the catalog at $\lambda_\star >2$, which is a typical value to reject single and pairs of galaxies and minimize the number of unrealistic and spurious detections. Additionally, we rejected detections falling into the first and last redshift bin which might be affected by border effects. 
The final catalog we produced contains 1678 detections in the range $0.08\leq z \leq3.7$ and with $S/N_{nocl}>6.0$. Despite AMICO has been widely used at $z<1$ \citep{bellagamba_amico_2018, maturi_amico_2019, maturi_minijpas_2023} and already tested \citep{euclid_collaboration_euclid_2019}, and applied to real data \citep{toni_amico-cosmos_2024} up to $z=2$, the detection of groups and clusters at even higher redshift, where they are still taking shape, is a quite unexplored regime, that we are interested in addressing with this work. The possibility to detect objects at $z>2$ with AMICO and the resulting sample derived from the application to the COSMOS-Web data are discussed in more detail in Sect. \ref{highz}.
For what concerns the sample at $z<2$, we compared the detections in the COSMOS-Web field with the previous COSMOS catalog constructed with AMICO \citep{toni_amico-cosmos_2024}. We performed a three-dimensional matching with maximum radial separation  d$r$ = 1 $Mpc/h$ and maximum redshift separation d$z$ = 0.05(1+$z$), equivalent to $\sim 200 \, cMpc$ at $z=2$, which are typical values for group/cluster matching. In the matching, we allowed multiple associations in order not to exclude potential cases of fragmentation or over-merging of detections. For this comparison, we worked on the common volume covered by the two catalogs, which is defined by the interval $0<z<2$ and by the smallest effective area, namely the one of the COSMOS-Web field. Therefore, we discarded masked detections\footnote{However, we allowed matching with masked detections if these fall within the tolerance of the matching.} from the previous AMICO-COSMOS catalog according to the visibility mask described in Sect. \ref{mask}. We found good correspondence between the two samples on the common volume, with a total number of 847 groups matched for the COSMOS-Web sample and 520 for the AMICO-COSMOS one. We explored the possibility for the difference in counts to signal fragmentation or over-merging. However, we found that the impact of fragmentation is marginal. Since we have used two different definitions of signal-to-noise as a lower limit to stop the detection process, we repeated the matching considering only COSMOS-Web detections with $S/N>3.0$ which was the minimum threshold of the AMICO-COSMOS catalog. This yielded a more consistent number of counts, with 400 matched detections for COSMOS-Web and 397 for AMICO-COSMOS. The percentage of matched detections for the AMICO-COSMOS catalog is around $\sim$97\% and increases to $\sim$98\% when considering only detections with $S/N>3.5$, which was identified as the threshold for selecting the most robust sample, based on the comparison between the detection performances in different photometric bands \citep[see][for details]{toni_amico-cosmos_2024}.

In Fig. \ref{statprop}, we show the redshift distribution of all the objects of this catalog and the cumulative signal-to-noise distribution, marking with the orange line the distributions for the detections with $\lambda_\star>10$, a value to select the rich-end tail of the distribution in $\lambda_\star$ and identify the richest objects in the catalog. In Fig. \ref{ls}, we plot the intrinsic richness, $\lambda_\star $ vs $z$ in three intervals of $S/N_{nocl}$. The trend of increasing intrinsic richness for increasing redshift observed up to $z \sim 2$ is expected, since the further we observe, the harder it is to detect poor and faint objects. At $z \gtrsim 2$, overdensities and numerous low-$\lambda_\star$ detections might be due to the fact that AMICO detects cores and substructures of clumpy extended protostructures rather than individual virialized clusters/groups. Additionally, the redshift interval $z \sim 2.4 -2.6$ is occupied in COSMOS by several large overdensities of galaxies known, for example, as Hyperion and Colossus protoclusters \citep[][see Table \ref{protocompilation} for further details and references]{cucciati_progeny_2018, lee_shadow_2016}. Figure \ref{fig:panelexamplpe} shows four examples of detections at different redshifts.
The final group catalog includes the columns described in Appendix \ref{appendixA} and it is accompanied by the list of member galaxies for each detection with membership and field probability as assigned by AMICO.
This galaxy membership probability is calculated after AMICO detects candidates in the 3D amplitude map, as briefly described in Sect. \ref{amico}. In particular, the probability of the $i$-th galaxy belonging to the $j$-th detection is expressed by

\begin{equation}\label{prob}
P_{i,j}= F_{i}\frac{A M_c(\boldsymbol{x}_i)p(z_j)}{A M_c(\boldsymbol{x}_i)p(z_j)+N} \, .
\end{equation}
Here, $F_i$ is the probability of belonging to the field, which is used as a scaling factor to consider any previous associations of the galaxy. The field probability assumes a value of 1 for each galaxy when the iterations start and decreases at each new association. The vector containing the galaxy properties as described in Sect. \ref{amico}, $\boldsymbol{x}_i$, contains here the sky position of the galaxy with respect to the group center. The noise and the value of the galaxy redshift probability distribution at the group redshift, $z_j$, are here expressed by $N$ and $p(z_j)$.
The galaxy member catalog and the membership probability for each galaxy associated with a detection are part of the output returned by AMICO. However, this membership association is by definition a model-dependent quantity. If one wants to study the properties of the clusters and groups themselves, such as the luminosity function or the density profile in a model-independent way, one needs to rely on a different method to identify possible members and field galaxies. For this reason, we included in our group catalog the possibility of retrieving statistical membership, as it was done for example by \cite{puddu_amico_2021} to study cluster luminosity functions. We proceeded as follows: in the membership catalog, we added a column named "\texttt{FLAG\_CYLINDER}", which we assigned the value “1” for all galaxies located within a cylinder centered in the center of each detection, with a radius of 0.5 $Mpc/h$ and depth $0.01(1+z)$; we then assigned the flag value “0” to all galaxies which do not fall in any of the cylinders. These flags help identify potential group and field galaxies without relying on the AMICO association probability. Besides this, we also created two three-dimensional maps of the volume covered by the group search, marking the effective area (taking into account only unmasked regions) inside or outside the cylinders, used to easily retrieve the group and field volumes in a given redshift slice and perform background subtraction.
This was done to make the membership catalog usable for galaxy population and group property studies.
However, when referring to group galaxies in this paper, we will make use of the membership probability directly returned by AMICO, selecting members as galaxies associated with $P > 50\%$, unless otherwise specified.

\subsection{Purity and completeness of the sample with SinFoniA}
We derived the purity and completeness of the sample by making use of data-driven mock catalogs, produced with the Selection Function extrActor \citep[SinFoniA;][]{maturi_amico_2019}. This method has already been applied to wide-field surveys such as KiDS \citep[][Maturi et al. in prep.]{maturi_amico_2019} and is currently part of the implementation of the Euclid pipeline. The idea is to generate mocks based on the input dataset used for the creation of our candidate catalog, which is already divided into field and group galaxies via the association probability returned by AMICO. After this, the AMICO algorithm is applied to the mock galaxy catalog in the same way as was done for the real data, and the list of generated mock groups is used as a "truth table" to study the resulting group catalog. Therefore, the reference catalog is matched with the results of the search and used to estimate the purity and completeness of the sample and the uncertainties on the retrieved properties of the groups.
The SinFoniA algorithm exploits a Monte Carlo approach to create realizations of the universe and generate realistic mocks directly from the data. Therefore, the generated mocks are not based on any model or assumption, and they are able to reproduce the complexity of the real data. This reduces the possibility of introducing biases due to the cosmological assumptions, which are, for example, behind numerical simulations. Here, we briefly describe the main steps of the approach we adopted for this specific application and the results of our comparison against realistic mocks. For a complete description of SinFoniA, we refer the reader to \cite{maturi_amico_2019}.

\begin{figure}
    \centering
    \includegraphics[width=1\linewidth]{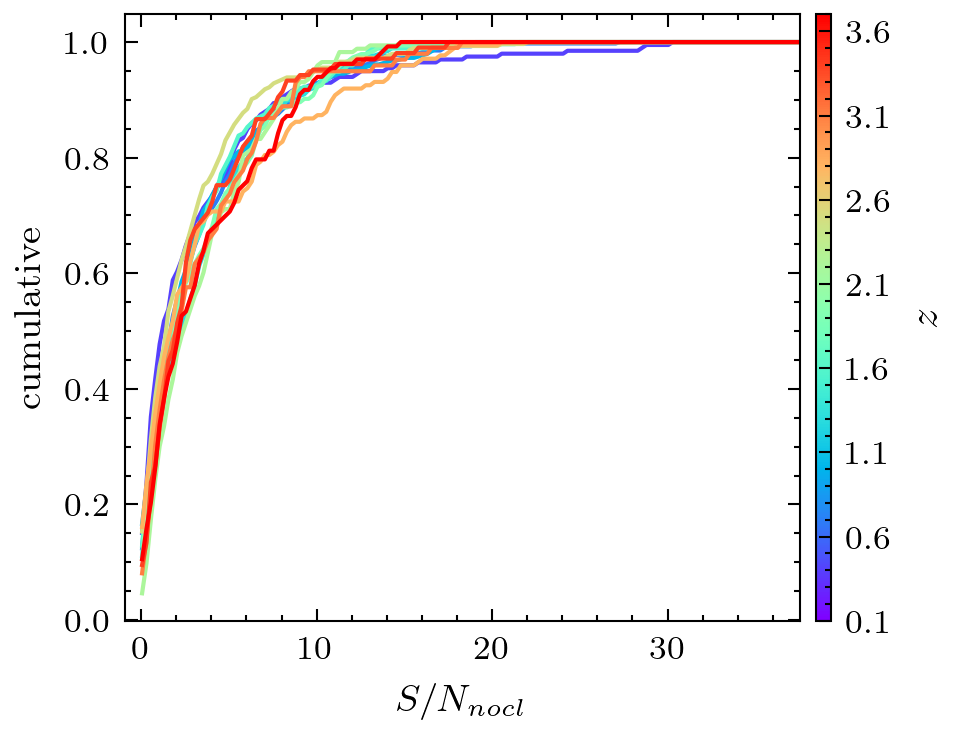}
    \caption{Cumulative distribution function (CDF) of $S/N_{nocl}$, in different bins of redshifts (color bar on the right), used to select detections the mocks are based on, instead of using a sharp signal-to-noise cut.}
    \label{fig:cdf}
\end{figure}

\noindent\textit{Mock catalog.} To create a data-driven mock catalog we proceeded as follows.
First, we derived the cumulative distribution function (CDF) of a group sample identical to the original catalog, but extending down to $S/N_{nocl}\sim 0$, and chose as reference ranking quantity the $S/N_{nocl}$ returned by AMICO. We did this in redshift bins to account for a possible redshift dependence. The CDF of this sample is shown in Fig. \ref{fig:cdf}, with different colors referring to different redshift bins. There is no significant redshift dependence for this application, proving once again the good quality of the photo-$z$ estimation. Then, we extracted the group detections using the CDF at their signal-to-noise ratios as the probability to be extracted, since this can be interpreted as the likelihood of the detection to correspond to a real group. This was done to avoid a sharp cut in $S/N_{nocl}$ when choosing which detections to consider. The field probability of galaxies is then recalculated by taking into account which detections have been rejected.
Secondly, we extracted the field galaxies by using the membership information. The probability of the galaxy belonging to the field ($F_i$ in Eq. \ref{prob}) is used as the probability for the galaxy to be extracted as part of the mock field. For this application, we kept the original position and redshift of the field galaxies to maintain the noise spatial correlation and so the main features of the large-scale structure.
Then, we generated the list of possible mock members. Once again, we extracted the possible members from the catalog created via rejection based on the CDF.

\begin{figure*}
    \centering
    \includegraphics[width=18.5cm]{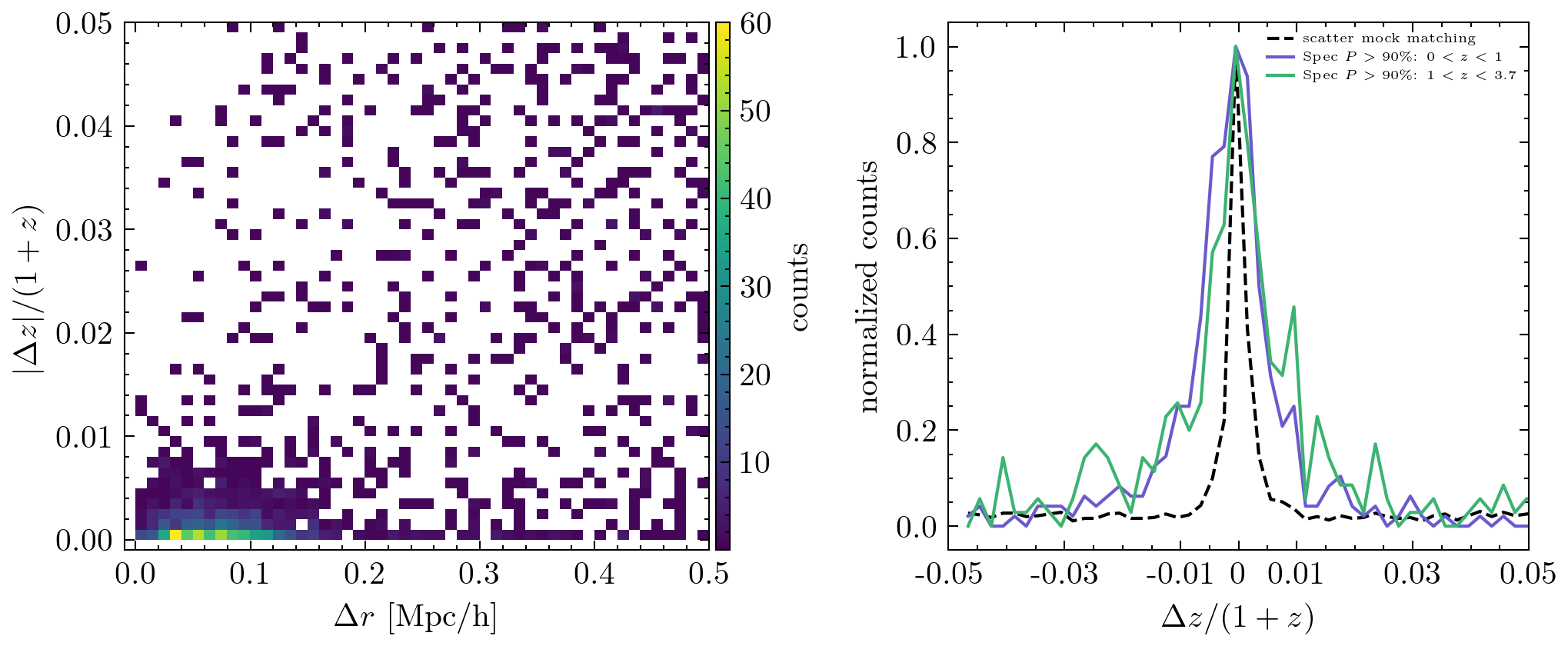}
    \caption{Left panel: Distribution of matched detections in the redshift scatter, $|\Delta z |/(1+z)$ - radial separation, $\Delta r$ [Mpc/$h$] plane for an initial matching with maximum separation d$z=0.05(1+z)$ and d$r=0.5$ Mpc/$h$. Most of the matched detections are concentrated in the rectangular area at $\Delta r < 0.2 $ Mpc/$h$ and at $|\Delta z |/(1+z) < 0.01$. Right panel: Distribution of the redshift scatter for the mock matching (dashed black line) compared to the scatter between group photo-$z$ and group spec-$z$ estimated as the mean of spec-members with $P>90\%$, in two intervals of redshift (solid purple and green lines). The distributions are normalized to 1.0 to better compare scatters.}
    \label{fig:scattermatch}
\end{figure*}
\begin{figure*}
    \centering
    \includegraphics[width=18.5cm]{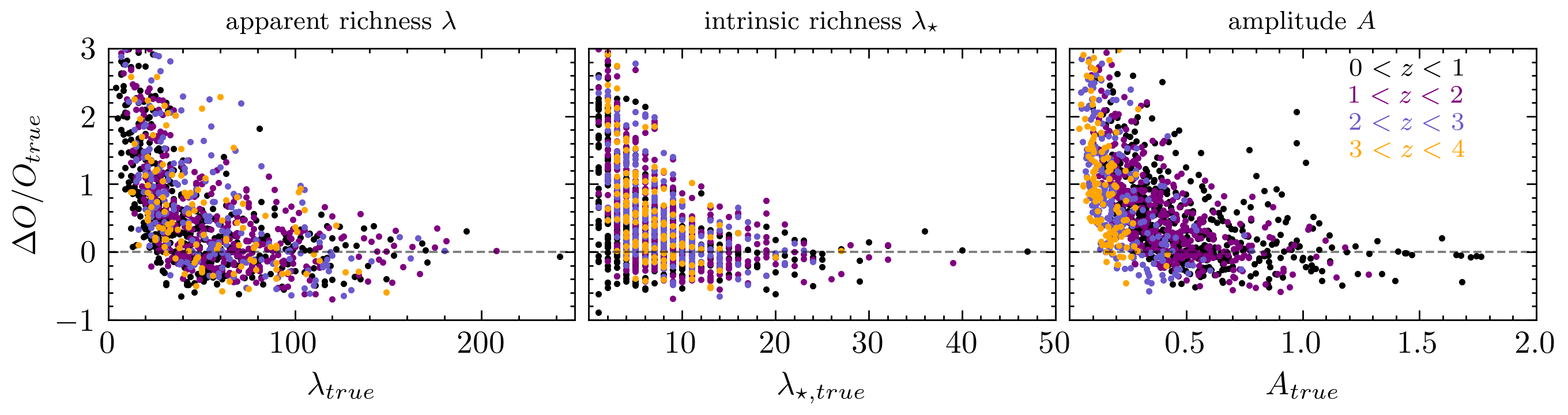}
    \caption{Relative scatter of the three observables ($O$, i.e., $\lambda$, $\lambda_\star$ and $A$, from left to right) between the detected observables, $O_{det}$, and the true observables as in the mocks, ($O_{true}$). The scatter is here expressed by $\Delta O = O_{det}-O_{true}$. Different colors mark different redshift bins as indicated in the plot on the right. Points with a scatter larger than 3.0 and the detection in Fig. \ref{fig:overlap_xray} (which is much richer than the rest of the sample) are not shown for better visualization of scatters and biases. }
    \label{fig:scatterobserv}
\end{figure*}

\begin{figure}
    \centering
    \includegraphics[width=1\linewidth]{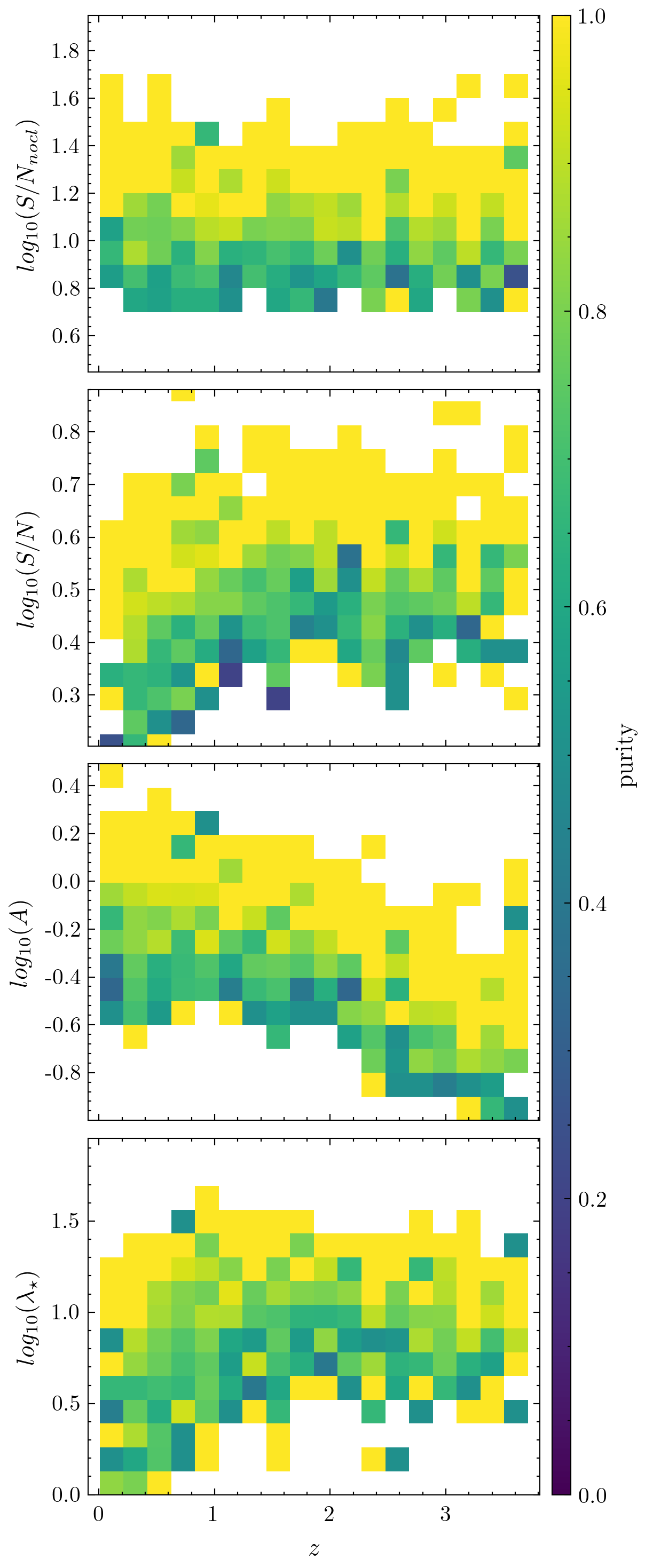}
    \caption{Purity of the group sample evaluated against the mock catalog produced with SinFoniA. The four panels show the redshift dependence of (non-cumulative) purity referred to four different reference detection properties: signal-to-noise ratio without cluster contribution, with cluster contribution, amplitude, and intrinsic richness (from top to bottom).}
    \label{fig:purity}
\end{figure}

All galaxies are then collected in bins of apparent richness and redshift so that the mock groups can be built by drawing from the full population in the corresponding bin. We used a bin resolution of $\Delta \lambda = 10$ and $\Delta z =0.01$ for apparent richness and redshift, respectively. With this approach, mock groups are built with members extracted from stacks of different original detections, using the membership probability ($P_{i,j}$ in Eq. \ref{prob}) as the probability of being extracted as members. However, this method tends to cancel out the intrinsic shape and orientation of clusters/groups. To restore their scatter in physical shape, we performed a coordinate transformation that does not affect density but introduces an ellipsoidal shape resembling that of dark matter halos in simulations \citep{despali_look_2017}.
Mock group position and redshift are randomly shuffled, with maximum displacement values of 0.25 $Mpc/h$ and 0.01 in redshift. These values were chosen in order to introduce a perturbation that is not large enough to alter the features and structures on large scales, namely to preserve the three-dimensional spatial correlation.
Finally, we ran AMICO on the generated mock catalog and used this as a reference to study the reliability of the original detections. Maturi et al. in prep. discuss the negligibility of the possible under-representation of small and blended objects in the mocks.\\

\noindent\textit{Matching procedure and group observables.} When assessing the quality of detections through matching with a reference, it is important to ensure the good quality of the matching procedure itself. To do so, we performed an initial matching with relatively loose tolerance, namely d$z$ = 0.05(1+$z$) and d$r$ = 0.5 $Mpc/h$, and used an a-priori sorting of the catalogs by richness to prioritize richer groups with respect to the more numerous poor groups. Then, we looked at the distribution of redshift and spatial separation between successfully matched detections, which is reported in the left panel of Fig. \ref{fig:scattermatch}. The majority of matched detections are concentrated at $\Delta r < 0.2 $ Mpc/$h$ and at $|\Delta z| /(1+z) < 0.01$, suggesting these might be more appropriate tolerance values to be chosen for the matching in order to discard those likely to be random matches (see Fig. \ref{fig:scattermatch}). We estimated the number of random matches by fitting the scatter distribution in redshift and position with a function given by the product of two Gaussian distributions, plus a constant that represents the value of the background consisting of random matches. We then repeated the same fitting with Cauchy distributions, which better represent the tails of the distribution. These tests showed that the background value is negligible with respect to the density inside the selected rectangle. 
We observed that typical scatter values in radial separation assume realistic values ($\Delta r = 0.2 $ Mpc/$h$) while for the redshift separation, the scatter distribution is very narrow, with most of the detections having their redshift matched within $\sim 0.005(1+z)$. In the right panel of Fig. \ref{fig:scattermatch}, we show the normalized distribution of scatter in redshift for the mock matching (dashed black line) and for the comparison between photo-$z$ and spec-$z$ of the detected groups with spectroscopic counterparts (solid purple and green lines, for two redshift intervals). For details about the spectroscopic counterpart assignment see Sect. \ref{bias}. The shown spec-$z$ values correspond to the mean spectroscopic redshift of members assigned with $P>90\%$. We also tested lower probability thresholds, but the scatter did not significantly change. It is visible that the scatter in redshift in the mocks is significantly smaller when compared to the spectroscopic scatter.

The matching against the true catalog allowed us to check for possible biases in the estimation of group observables, like the proxies of mass, amplitude, richness, and intrinsic richness. The relative scatter of the true and detected observable quantities is shown in Fig. \ref{fig:scatterobserv} for apparent richness, $\lambda$, intrinsic richness, $\lambda_\star$, and amplitude, $A$. No significant bias is visible in any of the observables and across the entire sample. As expected, the three proxies of mass are highly affected by the Malmquist bias, which is a selection effect and does not have to do with the detection method. This selection effect is significant for the smallest detections, as visible in Fig. \ref{fig:scatterobserv}, indicatively for $\lambda \lesssim 70$, for $\lambda_\star \lesssim 10$, and for $A \lesssim 0.5$, with a slight redshift dependence for the latter.

\begin{figure}
    \centering
    \includegraphics[width=1\linewidth]{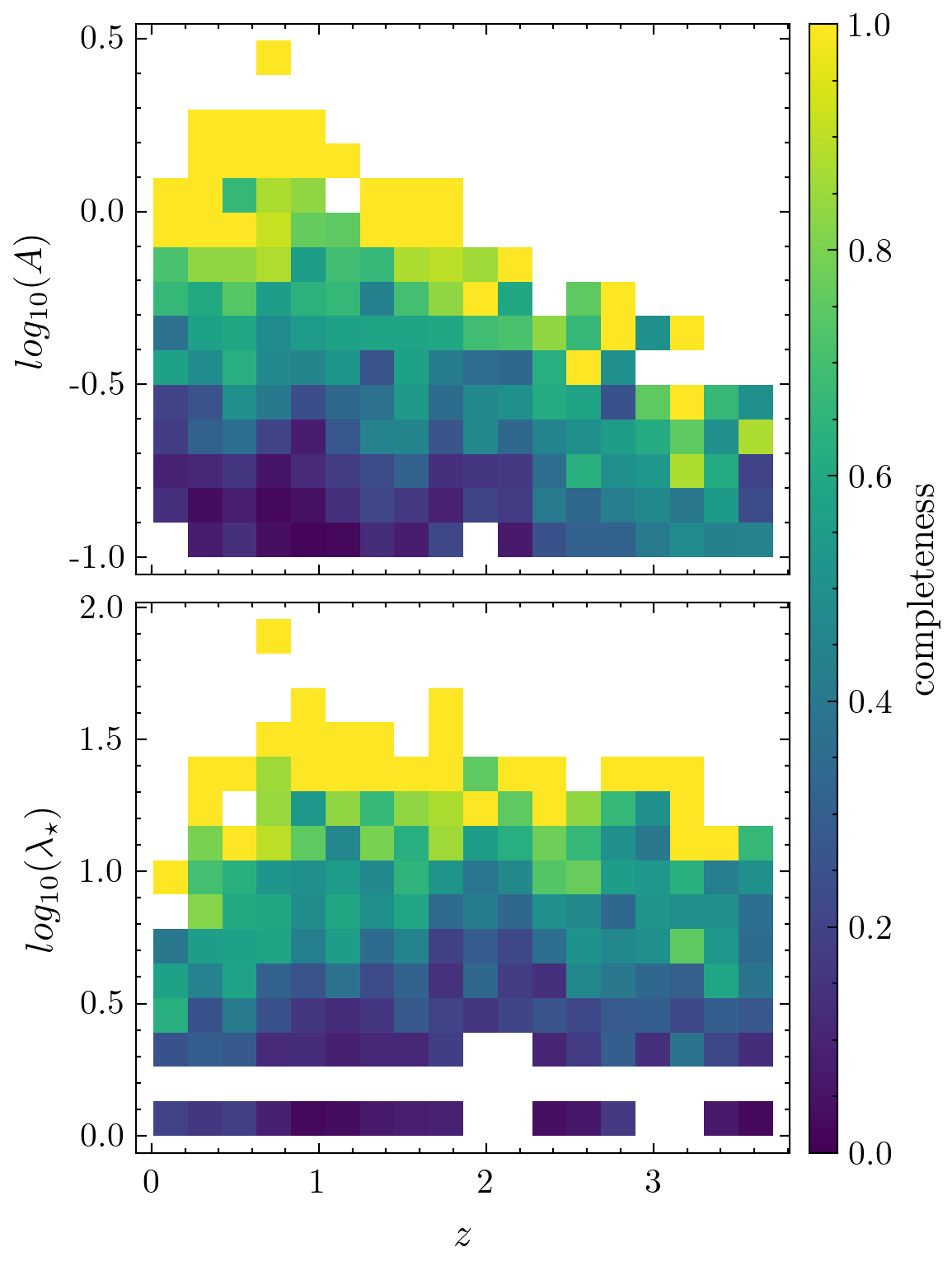}
    \caption{Completeness of the group sample evaluated against the mock catalog produced with SinFoniA. The two panels show the redshift dependence of completeness referred to two different proxies of mass returned by AMICO: amplitude (top) and intrinsic richness, $\lambda_\star$ (bottom).}
    \label{fig:comple}
\end{figure}

\begin{figure}
    \centering
    \includegraphics[width=1\linewidth]{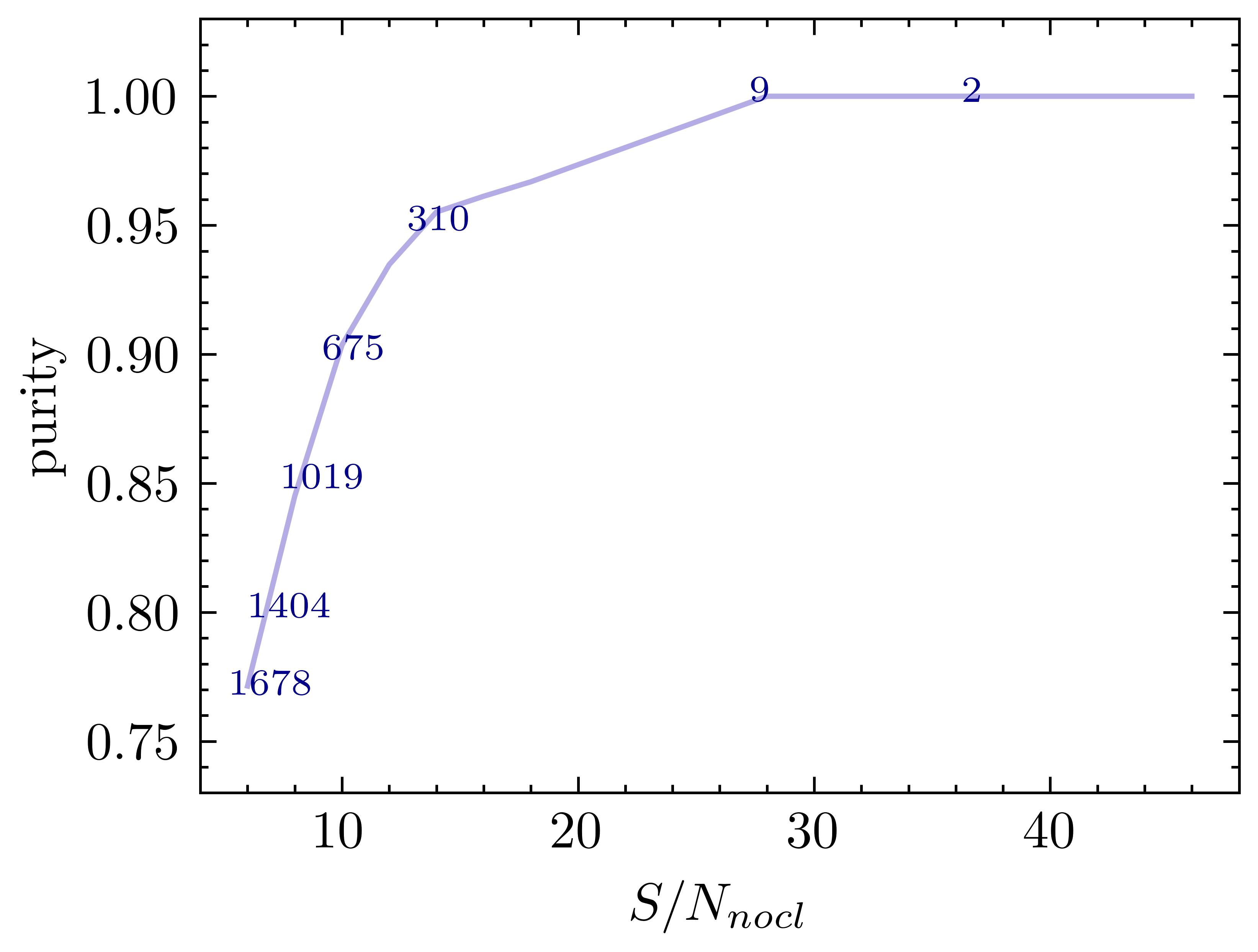}
    \caption{Relation between minimum $S/N_{nocl}$ and purity of the sample. In correspondence with some reference values of purity, we report the number of selected detections.}
    \label{fig:puritysnr}
\end{figure}

\noindent\textit{Purity and completeness.} The purity of a detected sample in reference to a true sample is defined as the ratio of the successfully matched detections to the total number of detections; the completeness is computed as the ratio of the successfully matched detections to the total number of true objects. We estimated the purity and the completeness of our sample against the mock catalog produced with SinFoniA in bins of detected redshift and detected candidate properties, the two signal-to-noise ratios, $S/N_{nocl}$ and $S/N$ and the two main mass-proxies returned by AMICO, the amplitude $A$ and the intrinsic richness, $\lambda_\star$, for the purity and the two mass proxies for the completeness. The values of purity and completeness at different redshifts and for the different properties are shown in Figs. \ref{fig:purity} and \ref{fig:comple}, respectively. As expected from the previous analyses of AMICO cluster and group samples (e.g., Maturi et al. in prep.), the $S/N_{nocl}$ value is the detection property with the best stability in redshift when analyzing the purity of the catalog. This means that this indication of signal-to-noise can be easily linked to a desired purity threshold. We found that, for this application, a cut at  $S/N_{nocl}\sim 10$ (selecting around 670 detections) would identify a $90\%$-pure group sample, while a cut at $S/N_{nocl}\sim 7$ (selecting around 1400 detections) would set purity at $80\%$. Figure \ref{fig:puritysnr} shows the signal-to-noise cut (horizontal axes) to be applied to this group sample for each desired level of purity (vertical axes). Concerning completeness, for this sample it resulted in being less redshift-dependent for the intrinsic richness (bottom panel of Fig. \ref{fig:comple}), with an 80\% completeness reached at around $\lambda_\star \sim 16$.

\subsection{Spectroscopic counterparts}\label{bias}

We associated spectroscopic redshifts with our detections, using the compilation of spectroscopic catalogs for the COSMOS field by \cite{khostovan_spec_compil_2025}. Public and private surveys included are, for instance, \cite{lilly_zcosmos_2007,lilly_zcosmos_2009,kartaltepe_rest-frame_2015,Hasinger_2018, kashino_fmos-cosmos_2019} and many others. Additional references can be found, for example, in \cite{gozaliasl_cosmos_2024}, Table 2. We selected sources with high-quality spectra, namely with a confidence level larger than 80\%. This yields a spectroscopic sample with more than 66,000 galaxies with spectroscopic redshift in the COSMOS field.
We assigned spectroscopic counterparts to our galaxy members identified by AMICO, via positional matching within 1$^{\prime \prime}$, which is the same separation used to assign counterparts from different surveys in the compilation \citep{khostovan_spec_compil_2025}. Thanks to the large spectroscopic coverage offered by the COSMOS field, we were able to assign a spec-$z$ to around 20\% of the group members assigned by AMICO (4300 members assigned with $P>50\%$ have a high-quality spectroscopic redshift). We defined the quality of the spectroscopic association to the group as $Q$, which is the ratio of the sum of the membership probabilities of the spectroscopic members to the number of spectroscopic members, $Q=\sum_i P_i/N_{gal}$. We found that 535 detections in total have at least three members with a spec-$z$. 948 detections have at least one galaxy member with spec-$z$ and $Q>70\%$. A total of 1075 detections have at least 3 spectroscopic members or an association with $Q>70\%$. Among these, around 80\% of the group candidates have the redshift assigned by AMICO that is consistent with $z_{spec}$. Here, $z_{spec}$ is defined as the mean spectroscopic redshift of the associated members. The consistency criterion is met when the relative difference between the two redshifts is less than 3\%, that is, $|\Delta z| /(1+z_{spec})<0.03$, which is around the largest $\sigma$ of the photo-$z$ uncertainty (see Sect. \ref{dataset}).
By assigning spectroscopic counterparts to our group members, we were able to exclude the presence of any significant general $z$-dependent bias between photo-$z$s and spec-$z$s, namely between AMICO redshift (based on galaxy photo-$z$s) and spectroscopic redshift of the group, estimated as the average redshift of the available spectroscopic members. This kind of bias was, for instance, found in the KiDS sample at low redshift and consequently corrected (Maturi et al. in prep.). The mean bias of the sample with more than 3 spectroscopic members is $\Delta z /(1+z_{spec}) = -0.011 \pm 0.051$.

However, even if a significant mean bias characterizing the sample is not visible, individual detections may show an inconsistency between the redshift assigned by AMICO and the one based on the available spectroscopic members. Therefore, we studied and marked with a dedicated flag all the detections that have a spectroscopic inconsistency of redshifts. 
For this analysis, we used only groups with at least 4 spectroscopic members or with spectroscopic quality $Q>95\%$. The flag is assigned to all detections that have relative redshift scatter,  $|\Delta z| /(1+z_{spec})>0.15$, which is the same limit chosen to compute outlier fractions for galaxies in the COSMOS-Web galaxy catalog \citep[see e.g.,][]{casey_cosmos-web_2023,shuntov_cosmos-web_2024}. Only 30 detections are affected by spectroscopic mismatch as just described. This confirms the already assessed good photo-$z$ quality of COSMOS-Web data used for this group search \citep{shuntov_cosmos-web_2024,arango-toro_history_2024}.
In this work, we also leveraged the availability of a large compilation of spec-$z$s to test the AMICO algorithm on a hybrid galaxy sample using also the spectroscopic redshifts, when available. This application is described in Appendix \ref{appendixB}.

\subsection{Detection flags}\label{qualityflags}
Besides flagging the detections in our catalog without spectroscopic members and those with spectroscopic members with redshift not compatible with the photometric one, we added useful flags to help filter the sample depending on the study that has to be performed with this catalog. 
Our flagging system is based on a list of base-2 flag bits referring to the following properties, ordered starting from the least severe:

\begin{enumerate}
    \item lack of spectroscopic members
    \item less than 3 arcmin from a border edge
    \setItemnumber{4}
    \item X-ray projection or proximity flag \footnote{In the same way as it was done in \cite{toni_amico-cosmos_2024} to clean the sample for the calibration of scaling relations, this flag marks detections within 0.02 deg from the center of a rich detection or lying on their extended X-ray emission.}
    \setItemnumber{8}
    \item masked fraction larger than 25\%
    \setItemnumber{16}
    \item central X-ray selected AGN  (obtained via matching with the X-ray sources in COSMOS-Web or COSMOS2020)
    \setItemnumber{32}
    \item spectroscopic mismatch (see Sect. \ref{bias})
    \setItemnumber{64}
    \item low intrinsic richness ($\lambda_\star<5$)
\end{enumerate}

\noindent In our detection catalog, we introduced the flag "\texttt{DETECTION\_FLAG}" which represents the sum of all these detection flag bits. Thus, if more than one criterion is present, the flags are summed together; for example, a group without spectroscopic members and $\lambda_\star=3$ will be flagged with 65.
According to this flagging system, around 67\% of the entire sample down to $S/N=6.0$ is flagged with a value smaller than 20, which is an example of how to select robust detections based on their individual properties.

\section{Candidates at z $\geq$ 2}\label{highz}
The depth of this catalog and the availability of high-quality photometric redshifts make this application ideal for testing the AMICO algorithm in the relatively unexplored regime of high-redshift groups, protogroups, and protoclusters.

\subsection{COSMOS protocluster compilation}
In order to benchmark our results in this challenging regime, we first collected literature about structures already known in the COSMOS field. We created a compilation of all currently known clusters, groups, protoclusters, or protocluster cores in the range $2.0 \lesssim z \lesssim 4.0$, which is the interval we are interested in for this comparison.  

The collected detections with relative references and sample properties, like position and redshift, are summarized in Table \ref{protocompilation}.
It should be noted that for this application of the AMICO algorithm to the COSMOS-Web field, we made use of a cluster model, as described in Sect. \ref{amico}, so we are relying on our knowledge of how galaxies are distributed in clusters and groups at low redshift, which might not be what we actually see at $z>1.5$. According to currently favored scenarios for cluster formation, present-day galaxy clusters and groups are the aftermath of the assembling, growing, and maturing of protocluster cores \citep[see e.g.,][]{shimakawa_mahalo_2018}.
In this high-$z$ group search, we expect the algorithm to detect cores or possibly virialized substructures of protoclusters which may extend for several tens of Mpcs. For this reason, to perform a consistent comparison of our high-$z$ sample with the known structures in the literature, we considered peaks and substructures inside extended protoclusters as individual objects. Just as an example, the 10 density peaks of Elentari \citep{forrest_elentarimassive_2023} and the 7 peaks of Hyperion \citep{cucciati_progeny_2018} are counted as 17 different objects in our protocluster compilation (see Table \ref{protocompilation}). In this compilation, we included not only objects detected as overdensities of photometric or spectroscopic redshifts \citep[e.g.,][]{chiang_discovery_2014, diener_proto-groups_2013,sarron_detectifz_2021}, but also discovered with several other approaches and methods, using, for example, the emission from distant galaxies or radio-galaxies \citep[e.g.,][]{geach_clustering_2012,castignani_cluster_2014,daddi_evidence_2022}, the mapping of Lyman $\alpha$ forest \citep[e.g.,][]{lee_ly_2014} and the group X-ray emission \citep[e.g.,][Gozaliasl et al. in prep.]{wang_discovery_2016}. Since this protocluster compilation might be used as a reference also in other studies beyond the comparison in this work, we included all detections in the COSMOS field, and not only in the COSMOS-Web portion of it. In case the object falls outside the COSMOS-Web field, we make that explicit in Table \ref{protocompilation}. Additionally, we did not perform internal matching to discard possible double detections, since it is not straightforward to define the spatial limits of protoclusters. However, if the discovery of a structure is generally attributed to more than one work, we report the corresponding references (see Col. 6 of Table \ref{protocompilation}). 

\subsection{Our candidates in the COSMOS-Web field}

\begin{figure*}[h]
   \centering
   \includegraphics[width=18cm]{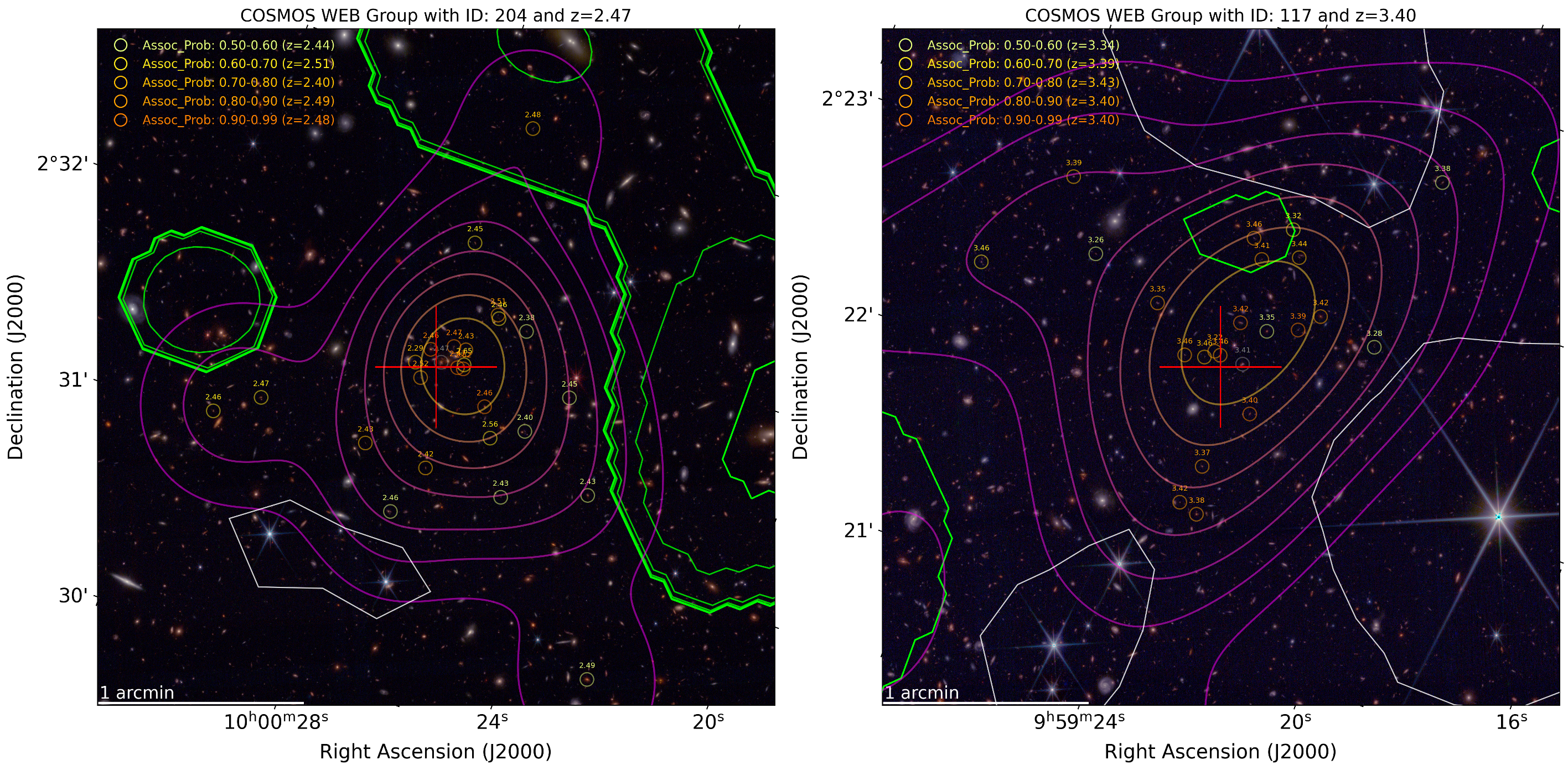}
      \caption{Two examples of high-$z$ detections not known in the literature. Stamps are JWST color-composite images, annotations are the same as described in the caption of Fig. \ref{fig:panelexamplpe}.  Left: Candidate at $z=2.47$ with $\lambda_\star \sim 11$. Right: Candidate at $z=3.40$ with $\lambda_\star \sim 25$. }
         \label{example1}
\end{figure*}

We ran a three-dimensional matching between our catalog and the compilation, in which we allow an AMICO detection to have more correspondences in the compilation, given that we did not discard objects potentially detected multiple times, as previously mentioned. The matching was run as follows: for each AMICO detection, we attributed a successful match with every other known object that lies within a cylinder of radius d$r$ = 1.0 $Mpc/h$ and redshift depth d$z$ = 0.05(1+$z$). These values are similar to scatter values chosen in literature for matching clusters/groups at very high redshift \citep[see e.g.,][]{sarron_detectifz_2021}. In this procedure, we used an a-priori sorting of the AMICO catalog by $(S/N)_{nocl}$.
In the last column of Table \ref{protocompilation}, the matched protoclusters are indicated with a checkmark. Whenever the reference detection is not a single object but a catalog or a structure with multiple peaks, we indicate next to the checkmark the percentage of matched peaks or detections over the total number of objects falling in the available area according to our visibility mask (described in Sect. \ref{mask}).
We successfully matched 205 of our groups to be compatible with detected protoclusters or substructures of protoclusters at $z \geq 2$ already known in literature, like the Hyperion protocluster structure \citep{cucciati_progeny_2018,casey_massive_2015,wang_discovery_2016}, Elentari \citep{forrest_elentarimassive_2023}, CC2.2 \citep{darvish_spectroscopic_2020}, and others.
We identified a total number of 316 new high-$z$ objects\footnote{During the revision phase of this paper, a catalog of overdensity peaks was published by \cite{hung_protocompilation_2025}. This catalog includes objects compatible with $\sim$ 49\% of our new detections. The overall matching rate is $\sim 93\%.$} with $2\leq z \leq 3.7$. These are interesting candidate protocluster cores and protogroups that do not match to any of the structures we collected from the literature in the field. Two examples of new high-$z$ detections are shown in Fig. \ref{example1}, one object is at $z=2.47$ and one at $z=3.40$.

\begin{figure}[h]
   \centering
   \includegraphics[width=9cm]{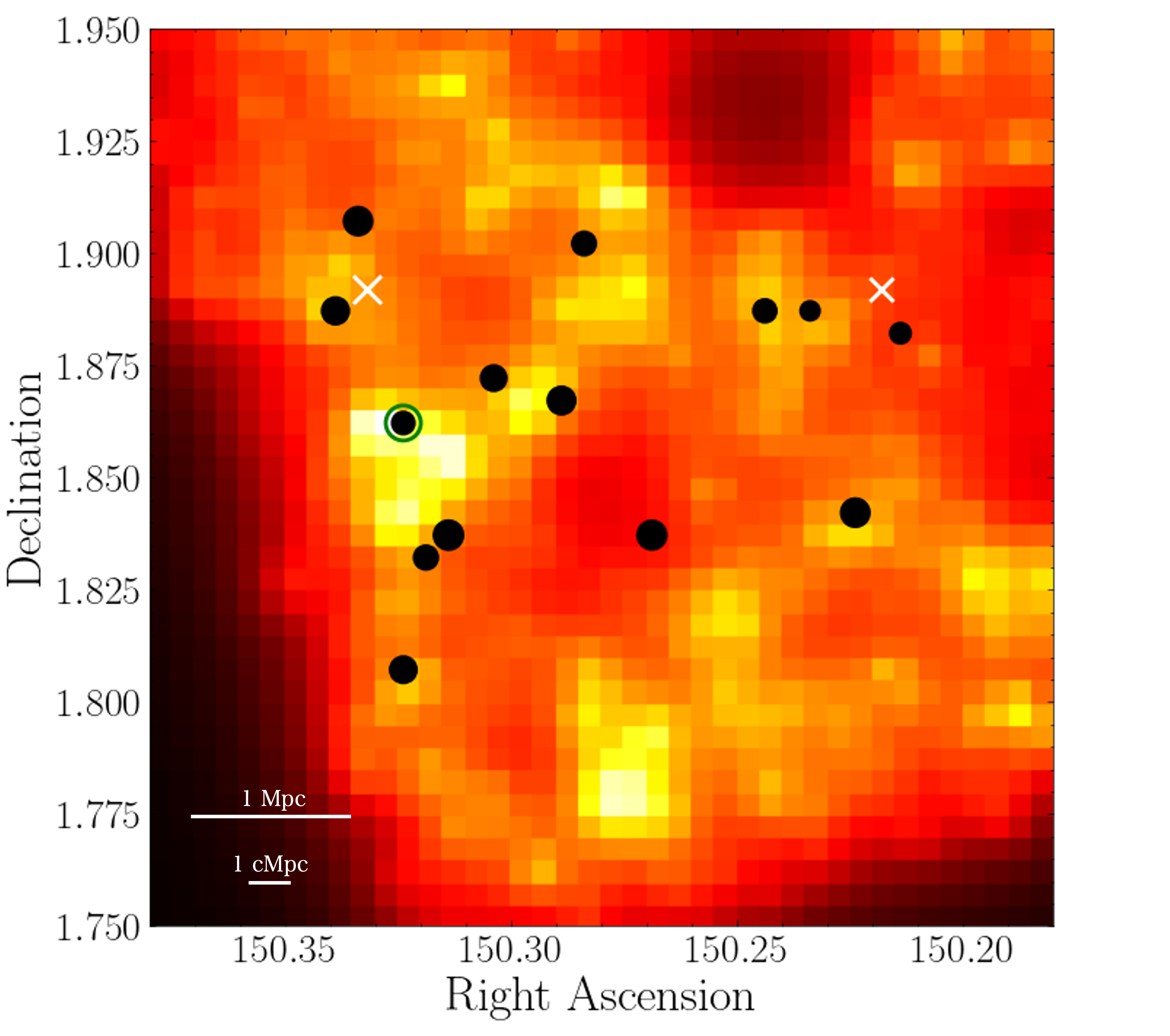}
      \caption{AmicOne protocluster candidate, consisting of fourteen cores (AMICO detections) with $z \in [2.5, 3.09]$ (black points) clustered together in the same large-scale structure, according to our clustering analysis. In the background, the amplitude map returned by AMICO at the redshift slice of the detection marked by the green circle, namely at $z=2.65$. Detections (black points) have different redshifts in the range, this is why they are not all lying on an amplitude peak. The amplitude peaks at their individual redshifts. The white crosses mark the position of the protogroups found by \cite{diener_proto-groups_2013}. The size of the points is proportional to the redshift.}
         \label{protocand}
\end{figure}

As mentioned above, at such high redshifts, we detected with AMICO groups that might be cores or density peaks of more extended protoclusters. For this reason, we explored the possibility of identifying protoclusters made of multiple AMICO detections by applying a 3D clustering algorithm. We treated our high-$z$ detections as points with 3D coordinates and performed a clustering analysis with the \texttt{sklearn} algorithm for Density-Based Spatial Clustering of Applications with Noise \citep[\texttt{DBSCAN};][]{ester_density-based_1996}. We used a minimum of two members per cluster, to allow any cluster size down to pairs of detections, and a clustering scale of 0.05 $deg$ as angular separation and $0.02(1+z)$ as maximum redshift separation. This analysis resulted in 111 potential large-scale protoclusters, with a maximum size of 14 detections in the same object. We show the largest of these structures in Fig. \ref{protocand}, to which we gave the name of AmicOne (pronounced \textipa{[a"mikone]} or \textipa{[a"mikw\textturnv n]}), the first and largest protocluster candidate detected with AMICO. This structure is formed by the clustering of 14 different cores, detected in the range $2.5\leq z\leq 3.09$, and it includes two groups consistent with those (IDs 37 and 47 in our compilation, in Table \ref{protocompilation}) detected by \citet{diener_proto-groups_2013}. The diametral size of the structure is about 20 cMpc on the plane of the sky at $z=2.65$ and it extends over a wide range of redshifts, equivalent to $\sim 600$ cMpc.

\begin{table*}[h]
      \caption[]{Compilation of protoclusters and high-redshift clusters and groups known in the literature for the COSMOS field.}
         \label{protocompilation}
     $$ 
         \begin{tabular}{c c c c c c c}
            \hline
            \hline
            \noalign{\smallskip}
             ID & Name & RA [deg] & DEC [deg] & $z$ & Reference(s) & Matched  \\
            \noalign{\smallskip}
            \hline
            \noalign{\smallskip}
            1 & ZFOURGE/ZFIRE & 150.094 & 2.251 & 2.095 & (1), (2) & \checkmark  \\
            \noalign{\smallskip}
            2 & G237  & 150.507 & 2.312 & 2.160 &(3)& \\
            \noalign{\smallskip}
            311 & - & 150.474 & 2.328 & 2.160 & (4) & \checkmark \\
            \noalign{\smallskip}
            315 & - & 150.208& 2.015 & 2.230 & (5) & \checkmark \\
            \noalign{\smallskip}
            3 & CC2.2 & 150.197 & 2.003 & 2.232  & (6)& \checkmark \\
            \noalign{\smallskip}
            314 & - & 150.060 & 2.200 & 2.300 & (7) &  \\
            \noalign{\smallskip}
            12-16 & COSTCO &[149.706, 150.129] & [2.024, 2.275] & [2.047, 2.391] & (8), (9), (10)& \checkmark \, (100\%)\\
            \noalign{\smallskip}
            312 & Colossus & 150.060 & 2.310 & 2.440 & (11), (7), (12), (13) &  \\
            \noalign{\smallskip}
            313 & - & 150.130 & 2.373 & 2.470 & (14) & \checkmark\\
            \noalign{\smallskip}
            310 & CL J1001+0220 &150.250& 2.333  & 2.506 & (15) &  \checkmark \\
            \noalign{\smallskip}
            5-11 & Hyperion & [149.958, 150.331] & [2.112, 2.404] & [2.423, 2.507]& (16)& \checkmark \, (100\%) \\
            \noalign{\smallskip}
            341-345 & LATIS1-D2 & [149.532, 150.347] & [1.988, 2.274] & [2.457, 2.685]& (17)& \checkmark \, (100\%) \\
            \noalign{\smallskip}
            91-266 & - & [149.409, 150.749] & [1.628, 2.793] & [2.000, 2.840] &(18)& \checkmark \, (82\%)\\
            \noalign{\smallskip}
            4 & QO-1000 & 150.043 & 1.694 & 2.770 & (19) & $\oslash$ \\
            \noalign{\smallskip}
            335-340 & - & [149.601, 150.620] & [1.593, 2.289] & [2.010, 2.900] & (20), (21) & \checkmark \, (100\%) \tablefootmark{$\ast$}\\
            \noalign{\smallskip}
            316 & VPC-1000 & 150.068& 1.967 & 2.900 & (22) & \checkmark \\
            \noalign{\smallskip}
            17-55 & - & [149.870, 150.588 ] & [1.766, 2.603] & [2.013, 2.957] & (23)& \checkmark \, (82\%)\\
            \noalign{\smallskip}
            267-297 & -  & [149.509, 150.693] & [1.707, 2.674] & [2.070, 3.080] & (24) & \checkmark \, (92\%) \\
            \noalign{\smallskip}
            334 & MAGAZ3NE & 150.429 & 2.506 & 3.120 & (25) & \checkmark\\
            \noalign{\smallskip}
            321-326 & - & [149.583, 150.346] & [1.967, 2.603] & [2.194, 3.295] & (26), (27) & \checkmark \, (80\%)\\
            \noalign{\smallskip}
            317-320 & CCPC & [149.980, 150.070] & [2.110, 2.280] & [2.098, 3.303] & (28) & \checkmark \, (100\%)\\
            \noalign{\smallskip}
            308-309 & MAGAZ3NE & [149.850, 150.117] & [2.333, 2.564]& [3.367, 3.380] & (29) & \checkmark \, (100\%)\\ 
            \noalign{\smallskip}
            298-307 & Elentari & [149.732, 150.424]  & [2.275, 2.583]& [3.248, 3.409]& (30) & \checkmark \, (100\%) \\
            \noalign{\smallskip}
            56-90 & - & [149.410, 150.616] & [1.543, 2.794] & [2.000, 3.500] &  (31)& \checkmark \,  (82\%) \\
            \noalign{\smallskip}
            327-333 & - & [149.706, 150.751] & [1.798, 2.762] & [2.323, 3.613] & (32) & $\star$\\
            \noalign{\smallskip}
            \hline
            \hline
         \end{tabular}
     $$ 
     \tablebib{
     (1)~\citet{spitler_first_2012}; (2) \citet{yuan_keckmosfire_2014}; (3) \citet{polletta_spectroscopic_2021}; (4) \citet{koyama_planck-selected_2021}; (5) \citet{geach_clustering_2012}; (6) \cite{darvish_spectroscopic_2020}; (7) \cite{lee_shadow_2016}; (8) \cite{ata_predicted_2022}; (9) \cite{dong_observational_2023}; (10) \cite{edward_stellar_2024}; (11) \cite{lee_ly_2014}; (12) \cite{diener_protocluster_2015}; (13) \cite{chiang_surveying_2015}; (14) \cite{casey_massive_2015}; (15) \cite{wang_discovery_2016}; (16) \cite{cucciati_progeny_2018}; (17) \cite{newman_protocl_2022}; (18) \cite{sarron_detectifz_2021}; (19) \cite{ito_cosmos2020_2023}; (20) \cite{castignani_cluster_2014}; (21) \cite{castignani_molecular_2019}; (22) \cite{cucciati_discovery_2014}; (23) \cite{diener_proto-groups_2013}; (24) \cite{chiang_discovery_2014}; (25) \cite{mcconachie_magaz3ne_2025}; (26) \cite{daddi_three_2021}; (27) \cite{daddi_evidence_2022}; (28) \cite{franck_candidate_2016}; (29) \cite{mcconachie_spectroscopic_2022}; (30) \cite{forrest_elentarimassive_2023}; (31) Gozaliasl et al., in prep.; (32) \cite{sillassen_noema_2024};
     }
     \tablefoot{The table shows arbitrary identification numbers in our compilation (ID), name of the protocluster or the survey, if available (Name), coordinate in degrees or range of coordinates (RA, DEC), redshift or range of redshifts ($z$), reference(s) and whether or not the objects are matched with one of our detection (and in which fraction if the reference is a catalog or list of objects).\\ 
     \tablefoottext{$\oslash$}{Outside the COSMOS-Web field; } 
     \tablefoottext{$\star$}{All outside the COSMOS-Web field except for one object, which is unmatched; } \tablefoottext{$\ast$}{All outside the COSMOS-Web field except for two objects, both matched.}}
   \end{table*}

\section{Conclusions}\label{conclusions}
We produced a deep galaxy group catalog based on the new COSMOS-Web photometric galaxy catalog down to $F150W < 27.3$. The detection procedure was performed with the AMICO algorithm, a widely tested cluster detector currently used in Euclid, KiDS, and other major surveys. The group search was performed over an effective area of 0.45 deg$^2$ and up to $z = 3.7$, covering the transition from protoclusters to virialized objects and their maturing phase. We detected 1678 groups with a signal-to-noise ratio ($S/N_{nocl}$) larger than 6.0. For each detection in our catalog, we provided information about position, redshift, signal-to-noise ratio, mass proxies (like amplitude $A$ and intrinsic richness, $\lambda_\star$), masked fraction, flags indicating detection features, like spectroscopic counterparts, presence of AGNs, and spectroscopic confirmation of redshift. In addition to the detection catalog, we also provide a list of members with their association probability and the information needed to estimate statistical membership. 

We evaluated the purity of our sample against realistic mocks generated with the SinFoniA algorithm. This method allows for the creation of mock catalogs, capturing the complexity of real data, without relying on any a-priori assumption. We found the relationship between purity and $S/N_{nocl}$, which is the detection property that yields the most redshift-independent relation with purity. This makes it possible to establish cuts of the original catalog based on the desired purity level. For example, we found that a cut at $S/N_{nocl}\sim 10$ would correspond to selecting a sample of around 670 objects with purity above $90\%$. 

We leveraged the successful application to COSMOS data up to $z=2$ presented in \cite{toni_amico-cosmos_2024} and the good quality of the photometric redshift and depth of the COSMOS-Web survey to create this deep catalog of galaxy groups and to explore the possibility of detecting protocluster cores with AMICO at $z \geq 2$. We successfully detected 316 new objects at $z \geq 2$ and 205 compatible with being part of protoclusters and high-$z$ groups known in the literature. To perform this comparative analysis, we created a compilation of known objects at $z>2$ in COSMOS and matched them with our detections. In total, our catalog contains 509 candidate groups and protocluster cores in the range $2 \leq z \leq 3.7$, of which around 400 are not isolated but lie within 111 Mpc-scale structures we found by performing a 3D clustering analysis with a minimum cluster size of 2, with the \texttt{DBSCAN} algorithm. Among these structures, the one including the highest number of substructures (14 cores) is a large protocluster candidate found at redshift $z \sim  2.5-3.0$ and we have assigned it the name AmicOne.

Such a deep and well-characterized sample of groups, extending over a wide range of richness and redshift, is an important resource for the study of group and cluster assembly and evolution, from high-$z$ cores to the massive objects we observe today. Additionally, it makes it possible to study several aspects of galaxy populations in different environments. For instance, it provides information about the numerous but relatively unexplored population of low-mass groups, and about galaxies in the outskirts of groups and clusters.

\begin{acknowledgements}
We acknowledge the contribution of the COSMOS collaboration, consisting of more than 200 scientists. More information about the COSMOS survey can be found at \url{https://cosmos.astro. caltech.edu/}. This work was made possible by utilizing the CANDIDE cluster at the Institut d’Astrophysique de Paris. The cluster was funded through grants from the PNCG, CNES, DIM-ACAV, the Euclid Consortium, and the Danish National Research Foundation Cosmic Dawn Center (DNRF140). It is maintained by Stephane Rouberol.
      LM acknowledges the financial contribution from the
PRIN-MUR 2022 20227RNLY3 grant “The concordance cosmological model:
stress-tests with galaxy clusters” supported by Next Generation EU and
from the grant ASI n. 2024-10-HH.0 “Attività scientifiche per la
missione Euclid – fase E”. This research was supported through the Visiting Scientist program of the International Space Science Institute (ISSI) in Bern. MF is supported by the European Union’s Horizon 2020 research and innovation programme under the Marie Sklodowska-Curie grant agreement No 101148925. GC acknowledges the support from the Next Generation EU funds within the National Recovery and Resilience Plan (PNRR), Mission 4 - Education and Research, Component 2 - From Research to Business (M4C2), Investment Line 3.1 - Strengthening and creation of Research Infrastructures, Project IR0000012 – “CTA+ - Cherenkov Telescope Array Plus”. ET acknowledges funding from the HTM (grant TK202), ETAg (grant PRG1006) and the EU Horizon Europe (EXCOSM, grant No. 101159513).
\end{acknowledgements}

\bibliographystyle{aa} 
\bibliography{references}

\begin{thebibliography}{135}
\expandafter\ifx\csname natexlab\endcsname\relax\def\natexlab#1{#1}\fi

\bibitem[{Alberts {et~al.}(2016)Alberts, Pope, Brodwin, Chung, Cybulski, Dey, Eisenhardt, Galametz, Gonzalez, Jannuzi, Stanford, Snyder, Stern, \& Zeimann}]{alberts_star_2016}
Alberts, S., Pope, A., Brodwin, M., {et~al.} 2016, \apj, 825, 72

\bibitem[{Allen {et~al.}(2011)Allen, Evrard, \& Mantz}]{allen_cosmological_2011}
Allen, S.~W., Evrard, A.~E., \& Mantz, A.~B. 2011, \araa, 49, 409

\bibitem[{Andreon {et~al.}(2014)Andreon, Newman, Trinchieri, Raichoor, Ellis, \& Treu}]{andreon_jkcs_2014}
Andreon, S., Newman, A.~B., Trinchieri, G., {et~al.} 2014, \aap, 565, A120

\bibitem[{{Arango-Toro} {et~al.}(2024){Arango-Toro}, Ilbert, Ciesla, Shuntov, Aufort, Mercier, Laigle, Franco, Bethermin, Le~Borgne, Dubois, McCracken, Paquereau, {Huertas-Company}, Kartaltepe, Casey, Akins, Allen, Andika, Brinch, Drakos, Faisst, Gozaliasl, Harish, Kaminsky, Koekemoer, Kokorev, Liu, Magdis, Martin, Moutard, Rhodes, Rich, Robertson, Sanders, Sheth, Talia, Toft, Tresse, Valentino, Vijayan, \& Weaver}]{arango-toro_history_2024}
{Arango-Toro}, R.~C., Ilbert, O., Ciesla, L., {et~al.} 2024, arXiv e-prints, arXiv.2410.05375

\bibitem[{Arnouts {et~al.}(2002)Arnouts, Moscardini, Vanzella, Colombi, Cristiani, Fontana, Giallongo, Matarrese, \& Saracco}]{arnouts_measuring_2002}
Arnouts, S., Moscardini, L., Vanzella, E., {et~al.} 2002, \mnras, 329, 355

\bibitem[{Ata {et~al.}(2022)Ata, Lee, Vecchia, Kitaura, Cucciati, Lemaux, Kashino, \& M{\"u}ller}]{ata_predicted_2022}
Ata, M., Lee, K.-G., Vecchia, C.~D., {et~al.} 2022, Nature Astronomy, 6, 857

\bibitem[{Balogh {et~al.}(2021)Balogh, {van der Burg}, Muzzin, Rudnick, Wilson, Webb, Biviano, Boak, Cerulo, Chan, Cooper, Gilbank, Gwyn, Lidman, Matharu, McGee, Old, {Pintos-Castro}, Reeves, Shipley, Vulcani, Yee, Alonso, Bellhouse, Cooke, Davidson, De~Lucia, Demarco, Drakos, Fillingham, Finoguenov, Forrest, Golledge, Jablonka, Lambas~Garcia, McNab, Muriel, Nantais, Noble, Parker, Petter, Poggianti, Townsend, Valotto, Webb, \& Zaritsky}]{balogh_gogreen_2021}
Balogh, M.~L., {van der Burg}, R. F.~J., Muzzin, A., {et~al.} 2021, \mnras, 500, 358

\bibitem[{Bamford {et~al.}(2009)Bamford, Nichol, Baldry, Land, Lintott, Schawinski, Slosar, Szalay, Thomas, Torki, Andreescu, Edmondson, Miller, Murray, Raddick, \& Vandenberg}]{bamford_galaxy_2009}
Bamford, S.~P., Nichol, R.~C., Baldry, I.~K., {et~al.} 2009, \mnras, 393, 1324

\bibitem[{Baxter {et~al.}(2023)Baxter, Cooper, Balogh, Rudnick, De~Lucia, Demarco, Finoguenov, Forrest, Muzzin, Reeves, Sarron, Vulcani, Wilson, \& Zaritsky}]{baxter_when_2023}
Baxter, D.~C., Cooper, M.~C., Balogh, M.~L., {et~al.} 2023, \mnras, 526, 3716

\bibitem[{Bellagamba {et~al.}(2011)Bellagamba, Maturi, Hamana, Meneghetti, Miyazaki, \& Moscardini}]{bellagamba_optimal_2011}
Bellagamba, F., Maturi, M., Hamana, T., {et~al.} 2011, \mnras, 413, 1145

\bibitem[{Bellagamba {et~al.}(2018)Bellagamba, Roncarelli, Maturi, \& Moscardini}]{bellagamba_amico_2018}
Bellagamba, F., Roncarelli, M., Maturi, M., \& Moscardini, L. 2018, \mnras, 473, 5221

\bibitem[{Bertin {et~al.}(2022)Bertin, Schefer, Apostolakos, {\'A}lvarez-Ayll{\'o}n, Dubath, \& K{\"u}mmel}]{bertin_sourcextractor_2022}
Bertin, E., Schefer, M., Apostolakos, N., {et~al.} 2022, Astrophys. Source Code Libr., ascl:2212.018

\bibitem[{Bianconi {et~al.}(2018)Bianconi, Smith, Haines, McGee, Finoguenov, \& Egami}]{bianconi_locuss_2018}
Bianconi, M., Smith, G.~P., Haines, C.~P., {et~al.} 2018, \mnras, 473, L79

\bibitem[{Bond {et~al.}(1996)Bond, Kofman, \& Pogosyan}]{bond_how_1996}
Bond, J.~R., Kofman, L., \& Pogosyan, D. 1996, Nature, 380, 603

\bibitem[{Boselli \& Gavazzi(2006)}]{boselli_environmental_2006}
Boselli, A. \& Gavazzi, G. 2006, \pasp, 118, 517

\bibitem[{Brodwin {et~al.}(2013)Brodwin, Stanford, Gonzalez, Zeimann, Snyder, Mancone, Pope, Eisenhardt, Stern, Alberts, Ashby, Brown, Chary, Dey, Galametz, Gettings, Jannuzi, Miller, Moustakas, \& Moustakas}]{brodwin_era_2013}
Brodwin, M., Stanford, S.~A., Gonzalez, A.~H., {et~al.} 2013, \apj, 779, 138

\bibitem[{Bruzual \& Charlot(2003)}]{bruzual_stellar_2003}
Bruzual, G. \& Charlot, S. 2003, \mnras, 344, 1000

\bibitem[{Capak {et~al.}(2007{\natexlab{a}})Capak, Abraham, Ellis, Mobasher, Scoville, Sheth, \& Koekemoer}]{capak_effects_2007}
Capak, P., Abraham, R.~G., Ellis, R.~S., {et~al.} 2007{\natexlab{a}}, \apjs, 172, 284

\bibitem[{Capak {et~al.}(2007{\natexlab{b}})Capak, Aussel, Ajiki, McCracken, Mobasher, Scoville, Shopbell, Taniguchi, Thompson, Tribiano, Sasaki, Blain, Brusa, Carilli, Comastri, Carollo, Cassata, Colbert, Ellis, Elvis, Giavalisco, Green, Guzzo, Hasinger, Ilbert, Impey, Jahnke, Kartaltepe, Kneib, Koda, Koekemoer, Komiyama, Leauthaud, Le~Fevre, Lilly, Liu, Massey, Miyazaki, Murayama, Nagao, Peacock, Pickles, Porciani, Renzini, Rhodes, Rich, Salvato, Sanders, Scarlata, Schiminovich, Schinnerer, Scodeggio, Sheth, Shioya, Tasca, Taylor, Yan, \& Zamorani}]{capak_first_2007}
Capak, P., Aussel, H., Ajiki, M., {et~al.} 2007{\natexlab{b}}, \apjs, 172, 99

\bibitem[{Casey {et~al.}(2022)Casey, Kartaltepe, \& {Cosmos-Web}}]{casey_cosmos-web_2022}
Casey, C., Kartaltepe, J., \& {Cosmos-Web}. 2022, 240, 203.02

\bibitem[{Casey {et~al.}(2015)Casey, Cooray, Capak, Fu, Kovac, Lilly, Sanders, Scoville, \& Treister}]{casey_massive_2015}
Casey, C.~M., Cooray, A., Capak, P., {et~al.} 2015, \apj, 808, L33

\bibitem[{Casey {et~al.}(2023)Casey, Kartaltepe, Drakos, Franco, Harish, Paquereau, Ilbert, Rose, Cox, Nightingale, Robertson, Silverman, Koekemoer, Massey, McCracken, Rhodes, Akins, Allen, Amvrosiadis, {Arango-Toro}, Bagley, Bongiorno, Capak, Champagne, Chartab, Ch{\'a}vez~Ortiz, Chworowsky, Cooke, Cooper, Darvish, Ding, Faisst, Finkelstein, Fujimoto, Gentile, Gillman, Gould, Gozaliasl, Hayward, He, Hemmati, Hirschmann, Jahnke, Jin, Khostovan, Kokorev, Lambrides, Laigle, Larson, Leung, Liu, Liaudat, Long, Magdis, Mahler, Mainieri, Manning, Maraston, Martin, McCleary, McKinney, McPartland, Mobasher, Pattnaik, Renzini, Rich, Sanders, Sattari, Scognamiglio, Scoville, Sheth, Shuntov, Sparre, Suzuki, Talia, Toft, Trakhtenbrot, Urry, Valentino, Vanderhoof, Vardoulaki, Weaver, Whitaker, Wilkins, Yang, \& Zavala}]{casey_cosmos-web_2023}
Casey, C.~M., Kartaltepe, J.~S., Drakos, N.~E., {et~al.} 2023, \apj, 954, 31

\bibitem[{Castignani {et~al.}(2014)Castignani, Chiaberge, Celotti, Norman, \& De~Zotti}]{castignani_cluster_2014}
Castignani, G., Chiaberge, M., Celotti, A., Norman, C., \& De~Zotti, G. 2014, \apj, 792, 114

\bibitem[{Castignani {et~al.}(2019)Castignani, Combes, Salom{\'e}, Benoist, Chiaberge, Freundlich, \& De~Zotti}]{castignani_molecular_2019}
Castignani, G., Combes, F., Salom{\'e}, P., {et~al.} 2019, \aap, 623, A48

\bibitem[{Catinella {et~al.}(2013)Catinella, Schiminovich, Cortese, Fabello, Hummels, Moran, Lemonias, Cooper, Wu, Heckman, \& Wang}]{catinella_galex_2013}
Catinella, B., Schiminovich, D., Cortese, L., {et~al.} 2013, \mnras, 436, 34

\bibitem[{Chartab {et~al.}(2021)Chartab, Mobasher, Shapley, Shivaei, Sanders, Coil, Kriek, Reddy, Siana, Freeman, Azadi, Barro, Fetherolf, Leung, Price, \& Zick}]{chartab_mosdef_2021}
Chartab, N., Mobasher, B., Shapley, A.~E., {et~al.} 2021, \apj, 908, 120

\bibitem[{Chiang {et~al.}(2014)Chiang, Overzier, \& Gebhardt}]{chiang_discovery_2014}
Chiang, Y.-K., Overzier, R., \& Gebhardt, K. 2014, \apj, 782, L3

\bibitem[{Chiang {et~al.}(2015)Chiang, Overzier, Gebhardt, Finkelstein, Chiang, Hill, Blanc, Drory, Chonis, Zeimann, Hagen, Schneider, Jogee, Ciardullo, \& Gronwall}]{chiang_surveying_2015}
Chiang, Y.-K., Overzier, R.~A., Gebhardt, K., {et~al.} 2015, \apj, 808, 37

\bibitem[{Civano {et~al.}(2016)Civano, Marchesi, Comastri, Urry, Elvis, Cappelluti, Puccetti, Brusa, Zamorani, Hasinger, Aldcroft, Alexander, Allevato, Brunner, Capak, Finoguenov, Fiore, Fruscione, Gilli, Glotfelty, Griffiths, Hao, Harrison, Jahnke, Kartaltepe, Karim, LaMassa, Lanzuisi, Miyaji, Ranalli, Salvato, Sargent, Scoville, Schawinski, Schinnerer, Silverman, Smolcic, Stern, Toft, Trakhtenbrot, Treister, \& Vignali}]{civano_chandra_2016}
Civano, F., Marchesi, S., Comastri, A., {et~al.} 2016, \apj, 819, 62

\bibitem[{Coupon(2018)}]{coupon_venice_2018}
Coupon, J. 2018, Astrophys. Source Code Libr., ascl:1802.002

\bibitem[{Coupon {et~al.}(2018)Coupon, Czakon, Bosch, Komiyama, Medezinski, Miyazaki, \& Oguri}]{coupon_bright-star_2018}
Coupon, J., Czakon, N., Bosch, J., {et~al.} 2018, \pasj, 70, S7

\bibitem[{Cucciati {et~al.}(2018)Cucciati, Lemaux, Zamorani, Le~F{\`e}vre, Tasca, Hathi, Lee, Bardelli, Cassata, Garilli, Le~Brun, Maccagni, Pentericci, Thomas, Vanzella, Zucca, Lubin, Amorin, Cassar{\`a}, Cimatti, Talia, Vergani, Koekemoer, Pforr, \& Salvato}]{cucciati_progeny_2018}
Cucciati, O., Lemaux, B.~C., Zamorani, G., {et~al.} 2018, \aap, 619, A49

\bibitem[{Cucciati {et~al.}(2014)Cucciati, Zamorani, Lemaux, Bardelli, Cimatti, Le~F{\`e}vre, Cassata, Garilli, Le~Brun, Maccagni, Pentericci, Tasca, Thomas, Vanzella, Zucca, Amorin, Capak, Cassar{\`a}, Castellano, Cuby, {de la Torre}, Durkalec, Fontana, Giavalisco, Grazian, Hathi, Ilbert, Moreau, Paltani, Ribeiro, Salvato, Schaerer, Scodeggio, Sommariva, Talia, Taniguchi, Tresse, Vergani, Wang, Charlot, Contini, Fotopoulou, {L{\'o}pez-Sanjuan}, Mellier, \& Scoville}]{cucciati_discovery_2014}
Cucciati, O., Zamorani, G., Lemaux, B.~C., {et~al.} 2014, \aap, 570, A16

\bibitem[{Daddi {et~al.}(2022)Daddi, Rich, Valentino, Jin, Delvecchio, Liu, Strazzullo, Neill, Gobat, Finoguenov, Bournaud, Elbaz, Kalita, O'Sullivan, \& Wang}]{daddi_evidence_2022}
Daddi, E., Rich, R.~M., Valentino, F., {et~al.} 2022, \apj, 926, L21

\bibitem[{Daddi {et~al.}(2021)Daddi, Valentino, Rich, Neill, Gronke, O'Sullivan, Elbaz, Bournaud, Finoguenov, Marchal, Delvecchio, Jin, Liu, Strazzullo, Calabro, Coogan, D'Eugenio, Gobat, Kalita, Laursen, Martin, Puglisi, Schinnerer, \& Wang}]{daddi_three_2021}
Daddi, E., Valentino, F., Rich, R.~M., {et~al.} 2021, \aap, 649, A78

\bibitem[{Darvish {et~al.}(2016)Darvish, Mobasher, Sobral, Rettura, Scoville, Faisst, \& Capak}]{darvish_effects_2016}
Darvish, B., Mobasher, B., Sobral, D., {et~al.} 2016, \apj, 825, 113

\bibitem[{Darvish {et~al.}(2020)Darvish, Scoville, Martin, Sobral, Mobasher, Rettura, Matthee, Capak, Chartab, Hemmati, Masters, Nayyeri, O'Sullivan, {Paulino-Afonso}, Sattari, Shahidi, Salvato, Lemaux, F{\`e}vre, \& Cucciati}]{darvish_spectroscopic_2020}
Darvish, B., Scoville, N.~Z., Martin, C., {et~al.} 2020, \apj, 892, 8

\bibitem[{Darvish {et~al.}(2014)Darvish, Sobral, Mobasher, Scoville, Best, Sales, \& Smail}]{darvish_cosmic_2014}
Darvish, B., Sobral, D., Mobasher, B., {et~al.} 2014, \apj, 796, 51

\bibitem[{Despali {et~al.}(2017)Despali, Giocoli, Bonamigo, Limousin, \& Tormen}]{despali_look_2017}
Despali, G., Giocoli, C., Bonamigo, M., Limousin, M., \& Tormen, G. 2017, \mnras, 466, 181

\bibitem[{Diener {et~al.}(2013)Diener, Lilly, Knobel, Zamorani, Lemson, Kampczyk, Scoville, Carollo, Contini, Kneib, Le~Fevre, Mainieri, Renzini, Scodeggio, Bardelli, Bolzonella, Bongiorno, Caputi, Cucciati, {de la Torre}, {de Ravel}, Franzetti, Garilli, Iovino, Kova{\v c}, Lamareille, Le~Borgne, Le~Brun, Maier, Mignoli, Pello, Peng, Perez~Montero, Presotto, Silverman, Tanaka, Tasca, Tresse, Vergani, Zucca, Bordoloi, Cappi, Cimatti, Coppa, Koekemoer, {L{\'o}pez-Sanjuan}, McCracken, Moresco, Nair, Pozzetti, \& Welikala}]{diener_proto-groups_2013}
Diener, C., Lilly, S.~J., Knobel, C., {et~al.} 2013, \apj, 765, 109

\bibitem[{Diener {et~al.}(2015)Diener, Lilly, Ledoux, Zamorani, Bolzonella, Murphy, Capak, Ilbert, \& McCracken}]{diener_protocluster_2015}
Diener, C., Lilly, S.~J., Ledoux, C., {et~al.} 2015, \apj, 802, 31

\bibitem[{Dong {et~al.}(2023)Dong, Lee, Ata, Horowitz, \& Momose}]{dong_observational_2023}
Dong, C., Lee, K.-G., Ata, M., Horowitz, B., \& Momose, R. 2023, \apj, 945, L28

\bibitem[{Dressler(1980)}]{dressler_galaxy_1980}
Dressler, A. 1980, \apj, 236, 351

\bibitem[{Edward {et~al.}(2024)Edward, Balogh, Bah{\'e}, Cooper, Hatch, Marchioni, Muzzin, Noble, Rudnick, Vulcani, Wilson, De~Lucia, Demarco, Forrest, Hirschmann, Castignani, Cerulo, Finn, Hewitt, Jablonka, Kodama, Maurogordato, Nantais, \& Xie}]{edward_stellar_2024}
Edward, A.~H., Balogh, M.~L., Bah{\'e}, Y.~M., {et~al.} 2024, \mnras, 527, 8598

\bibitem[{Eke {et~al.}(2004)Eke, Baugh, Cole, Frenk, Norberg, Peacock, Baldry, {Bland-Hawthorn}, Bridges, Cannon, Colless, Collins, Couch, Dalton, {de Propris}, Driver, Efstathiou, Ellis, Glazebrook, Jackson, Lahav, Lewis, Lumsden, Maddox, Madgwick, Peterson, Sutherland, \& Taylor}]{eke_galaxy_2004}
Eke, V.~R., Baugh, C.~M., Cole, S., {et~al.} 2004, \mnras, 348, 866

\bibitem[{Ester {et~al.}(1996)Ester, Kriegel, Sander, \& Xu}]{ester_density-based_1996}
Ester, M., Kriegel, H.-P., Sander, J., \& Xu, X. 1996, A {{Density-Based Algorithm}} for {{Discovering Clusters}} in {{Large Spatial Databases}} with {{Noise}}, 226--331

\bibitem[{{Euclid Collaboration} {et~al.}(2019){Euclid Collaboration}, Adam, Vannier, Maurogordato, Biviano, Adami, Ascaso, Bellagamba, Benoist, Cappi, {D{\'i}az-S{\'a}nchez}, Durret, Farrens, Gonzalez, Iovino, Licitra, Maturi, Mei, Merson, Munari, Pell{\'o}, Ricci, Rocci, Roncarelli, Sarron, Amoura, Andreon, Apostolakos, Arnaud, Bardelli, Bartlett, Baugh, Borgani, Brodwin, Castander, Castignani, Cucciati, De~Lucia, Dubath, Fosalba, Giocoli, Hoekstra, Mamon, Melin, Moscardini, Paltani, Radovich, Sartoris, Schultheis, Sereno, Weller, Burigana, Carvalho, Corcione, {Kurki-Suonio}, Lilje, Sirri, {Toledo-Moreo}, \& Zamorani}]{euclid_collaboration_euclid_2019}
{Euclid Collaboration}, Adam, R., Vannier, M., {et~al.} 2019, \aap, 627, A23

\bibitem[{Finoguenov {et~al.}(2007)Finoguenov, Guzzo, Hasinger, Scoville, Aussel, B{\"o}hringer, Brusa, Capak, Cappelluti, Comastri, Giodini, Griffiths, Impey, Koekemoer, Kneib, Leauthaud, Le~F{\`e}vre, Lilly, Mainieri, Massey, McCracken, Mobasher, Murayama, Peacock, Sakelliou, Schinnerer, Silverman, Smol{\v c}i{\'c}, Taniguchi, Tasca, Taylor, Trump, \& Zamorani}]{finoguenov_xmm-newton_2007}
Finoguenov, A., Guzzo, L., Hasinger, G., {et~al.} 2007, \apjs, 172, 182

\bibitem[{Forrest {et~al.}(2023)Forrest, Lemaux, Shah, Staab, McConachie, Cucciati, Gal, Hung, Lubin, Cassar{\`a}, Cassata, Chang, Cooper, Decarli, Gomez, Gururajan, Hathi, Kashino, Marchesini, Marsan, McDonald, Muzzin, Shen, Urbano~Stawinski, Talia, Vergani, Wilson, \& Zamorani}]{forrest_elentarimassive_2023}
Forrest, B., Lemaux, B.~C., Shah, E., {et~al.} 2023, \mnras, 526, L56

\bibitem[{Franck \& McGaugh(2016)}]{franck_candidate_2016}
Franck, J.~R. \& McGaugh, S.~S. 2016, \apj, 833, 15

\bibitem[{{Gaia Collaboration} {et~al.}(2018){Gaia Collaboration}, Brown, Vallenari, Prusti, {de Bruijne}, Babusiaux, {Bailer-Jones}, Biermann, Evans, Eyer, Jansen, Jordi, Klioner, Lammers, Lindegren, Luri, Mignard, Panem, Pourbaix, Randich, Sartoretti, Siddiqui, Soubiran, {van Leeuwen}, Walton, Arenou, Bastian, Cropper, Drimmel, Katz, Lattanzi, Bakker, Cacciari, Casta{\~n}eda, Chaoul, Cheek, De~Angeli, Fabricius, Guerra, Holl, Masana, Messineo, Mowlavi, Nienartowicz, Panuzzo, Portell, Riello, Seabroke, Tanga, Th{\'e}venin, {Gracia-Abril}, Comoretto, {Garcia-Reinaldos}, Teyssier, Altmann, Andrae, Audard, {Bellas-Velidis}, Benson, Berthier, Blomme, Burgess, Busso, Carry, Cellino, Clementini, Clotet, Creevey, Davidson, De~Ridder, Delchambre, Dell'Oro, Ducourant, {Fern{\'a}ndez-Hern{\'a}ndez}, Fouesneau, Fr{\'e}mat, Galluccio, {Garc{\'i}a-Torres}, {Gonz{\'a}lez-N{\'u}{\~n}ez}, {Gonz{\'a}lez-Vidal}, Gosset, Guy, Halbwachs, Hambly, Harrison, Hern{\'a}ndez, Hestroffer, Hodgkin, Hutton, Jasniewicz,
  {Jean-Antoine-Piccolo}, Jordan, Korn, {Krone-Martins}, Lanzafame, Lebzelter, L{\"o}ffler, Manteiga, Marrese, {Mart{\'i}n-Fleitas}, Moitinho, Mora, Muinonen, Osinde, Pancino, Pauwels, Petit, {Recio-Blanco}, Richards, Rimoldini, Robin, Sarro, Siopis, Smith, Sozzetti, S{\"u}veges, Torra, {van Reeven}, Abbas, Abreu~Aramburu, Accart, Aerts, Altavilla, {\'A}lvarez, Alvarez, Alves, Anderson, Andrei, Anglada~Varela, Antiche, Antoja, Arcay, Astraatmadja, Bach, Baker, {Balaguer-N{\'u}{\~n}ez}, Balm, Barache, Barata, Barbato, Barblan, Barklem, Barrado, Barros, Barstow, Bartholom{\'e}~Mu{\~n}oz, Bassilana, Becciani, Bellazzini, Berihuete, Bertone, Bianchi, Bienaym{\'e}, {Blanco-Cuaresma}, Boch, Boeche, Bombrun, Borrachero, Bossini, Bouquillon, Bourda, Bragaglia, Bramante, Breddels, Bressan, Brouillet, Br{\"u}semeister, Brugaletta, Bucciarelli, Burlacu, Busonero, Butkevich, Buzzi, Caffau, Cancelliere, Cannizzaro, {Cantat-Gaudin}, Carballo, Carlucci, Carrasco, Casamiquela, Castellani, {Castro-Ginard}, Charlot, Chemin,
  Chiavassa, Cocozza, Costigan, Cowell, Crifo, Crosta, Crowley, Cuypers, Dafonte, Damerdji, Dapergolas, David, David, {de Laverny}, De~Luise, De~March, {de Martino}, {de Souza}, {de Torres}, Debosscher, {del Pozo}, Delbo, Delgado, Delgado, Di~Matteo, Diakite, Diener, Distefano, Dolding, Drazinos, Dur{\'a}n, Edvardsson, Enke, Eriksson, Esquej, Eynard~Bontemps, Fabre, Fabrizio, Faigler, Falc{\~a}o, Farr{\`a}s~Casas, Federici, Fedorets, Fernique, Figueras, Filippi, Findeisen, Fonti, Fraile, Fraser, Fr{\'e}zouls, Gai, Galleti, Garabato, {Garc{\'i}a-Sedano}, Garofalo, Garralda, Gavel, Gavras, Gerssen, Geyer, Giacobbe, Gilmore, Girona, Giuffrida, Glass, Gomes, Granvik, Gueguen, Guerrier, Guiraud, {Guti{\'e}rrez-S{\'a}nchez}, Haigron, Hatzidimitriou, Hauser, Haywood, Heiter, Helmi, Heu, Hilger, Hobbs, Hofmann, Holland, Huckle, Hypki, Icardi, Jan{\ss}en, {Jevardat de Fombelle}, Jonker, Juh{\'a}sz, Julbe, Karampelas, Kewley, Klar, Kochoska, Kohley, Kolenberg, Kontizas, Kontizas, Koposov, Kordopatis,
  {Kostrzewa-Rutkowska}, Koubsky, Lambert, Lanza, Lasne, Lavigne, Le~Fustec, {Le Poncin-Lafitte}, Lebreton, Leccia, Leclerc, {Lecoeur-Taibi}, Lenhardt, Leroux, Liao, Licata, Lindstr{\o}m, Lister, Livanou, Lobel, L{\'o}pez, Managau, Mann, Mantelet, Marchal, Marchant, Marconi, Marinoni, Marschalk{\'o}, Marshall, Martino, Marton, Mary, Massari, Matijevi{\v c}, Mazeh, McMillan, Messina, Michalik, Millar, Molina, Molinaro, Moln{\'a}r, Montegriffo, Mor, Morbidelli, Morel, Morris, Mulone, Muraveva, Musella, Nelemans, Nicastro, Noval, O'Mullane, Ord{\'e}novic, {Ord{\'o}{\~n}ez-Blanco}, Osborne, Pagani, Pagano, Pailler, Palacin, Palaversa, Panahi, Pawlak, Piersimoni, Pineau, Plachy, Plum, Poggio, Poujoulet, Pr{\v s}a, Pulone, Racero, Ragaini, Rambaux, {Ramos-Lerate}, Regibo, Reyl{\'e}, Riclet, Ripepi, Riva, Rivard, Rixon, Roegiers, Roelens, {Romero-G{\'o}mez}, Rowell, Royer, {Ruiz-Dern}, Sadowski, Sagrist{\`a}~Sell{\'e}s, Sahlmann, Salgado, Salguero, Sanna, {Santana-Ros}, Sarasso, Savietto, Schultheis, Sciacca, Segol,
  Segovia, S{\'e}gransan, Shih, Siltala, Silva, Smart, Smith, Solano, Solitro, Sordo, Soria~Nieto, Souchay, Spagna, Spoto, Stampa, Steele, Steidelm{\"u}ller, Stephenson, Stoev, Suess, Surdej, Szabados, {Szegedi-Elek}, Tapiador, Taris, Tauran, Taylor, Teixeira, Terrett, Teyssandier, Thuillot, Titarenko, Torra~Clotet, Turon, Ulla, Utrilla, Uzzi, Vaillant, Valentini, Valette, {van Elteren}, Van~Hemelryck, {van Leeuwen}, Vaschetto, Vecchiato, Veljanoski, Viala, Vicente, Vogt, {von Essen}, Voss, Votruba, Voutsinas, Walmsley, Weiler, Wertz, Wevers, Wyrzykowski, Yoldas, {\v Z}erjal, Ziaeepour, Zorec, Zschocke, Zucker, Zurbach, \& Zwitter}]{gaia_collaboration_gaia_2018}
{Gaia Collaboration}, Brown, A. G.~A., Vallenari, A., {et~al.} 2018, \aap, 616, A1

\bibitem[{Gaspari {et~al.}(2011)Gaspari, Brighenti, D'Ercole, \& Melioli}]{gaspari_agn_2011}
Gaspari, M., Brighenti, F., D'Ercole, A., \& Melioli, C. 2011, \mnras, 415, 1549

\bibitem[{Geach {et~al.}(2012)Geach, Sobral, Hickox, Wake, Smail, Best, Baugh, \& Stott}]{geach_clustering_2012}
Geach, J.~E., Sobral, D., Hickox, R.~C., {et~al.} 2012, \mnras, 426, 679

\bibitem[{George {et~al.}(2012)George, Leauthaud, Bundy, Finoguenov, Ma, Rykoff, Tinker, Wechsler, Massey, \& Mei}]{george_galaxies_2012}
George, M.~R., Leauthaud, A., Bundy, K., {et~al.} 2012, \apj, 757, 2

\bibitem[{George {et~al.}(2011)George, Leauthaud, Bundy, Finoguenov, Tinker, Lin, Mei, Kneib, Aussel, Behroozi, Busha, Capak, Coccato, Covone, Faure, Fiorenza, Ilbert, Le~Floc'h, Koekemoer, Tanaka, Wechsler, \& Wolk}]{george_galaxies_2011}
George, M.~R., Leauthaud, A., Bundy, K., {et~al.} 2011, \apj, 742, 125

\bibitem[{Giodini {et~al.}(2009)Giodini, Pierini, Finoguenov, Pratt, Boehringer, Leauthaud, Guzzo, Aussel, Bolzonella, Capak, Elvis, Hasinger, Ilbert, Kartaltepe, Koekemoer, Lilly, Massey, McCracken, Rhodes, Salvato, Sanders, Scoville, Sasaki, Smolcic, Taniguchi, Thompson, \& {COSMOS Collaboration}}]{giodini_stellar_2009}
Giodini, S., Pierini, D., Finoguenov, A., {et~al.} 2009, \apj, 703, 982

\bibitem[{Gozaliasl {et~al.}(2024)Gozaliasl, Finoguenov, Babul, Ilbert, Sargent, Vardoulaki, Faisst, Liu, Shuntov, Cooper, Dolag, Toft, Magdis, Toni, Mobasher, Barr{\'e}, Cui, \& Rennehan}]{gozaliasl_cosmos_2024}
Gozaliasl, G., Finoguenov, A., Babul, A., {et~al.} 2024, \aap, 690, A315

\bibitem[{Gozaliasl {et~al.}(2019)Gozaliasl, Finoguenov, Tanaka, Dolag, Montanari, Kirkpatrick, Vardoulaki, Khosroshahi, Salvato, Laigle, McCracken, Ilbert, Cappelluti, Daddi, Hasinger, Capak, Scoville, Toft, Civano, Griffiths, Balogh, Li, Ahoranta, Mei, Iovino, Henriques, \& Erfanianfar}]{gozaliasl_chandra_2019}
Gozaliasl, G., Finoguenov, A., Tanaka, M., {et~al.} 2019, \mnras, 483, 3545

\bibitem[{Gunn \& Gott(1972)}]{gunn_infall_1972}
Gunn, J.~E. \& Gott, III, J.~R. 1972, \apj, 176, 1

\bibitem[{{Hasinger} {et~al.}(2018){Hasinger}, {Capak}, {Salvato}, {Barger}, {Cowie}, {Faisst}, {Hemmati}, {Kakazu}, {Kartaltepe}, {Masters}, {Mobasher}, {Nayyeri}, {Sanders}, {Scoville}, {Suh}, {Steinhardt}, \& {Yang}}]{Hasinger_2018}
{Hasinger}, G., {Capak}, P., {Salvato}, M., {et~al.} 2018, \apj, 858, 77

\bibitem[{Hasinger {et~al.}(2007)Hasinger, Cappelluti, Brunner, Brusa, Comastri, Elvis, Finoguenov, Fiore, Franceschini, Gilli, Griffiths, Lehmann, Mainieri, Matt, Matute, Miyaji, Molendi, Paltani, Sanders, Scoville, Tresse, Urry, Vettolani, \& Zamorani}]{hasinger_xmm-newton_2007}
Hasinger, G., Cappelluti, N., Brunner, H., {et~al.} 2007, \apjs, 172, 29

\bibitem[{Hausman \& Ostriker(1978)}]{hausman_galactic_1978}
Hausman, M.~A. \& Ostriker, J.~P. 1978, \apj, 224, 320

\bibitem[{Hennig {et~al.}(2017)Hennig, Mohr, Zenteno, Desai, Dietrich, Bocquet, Strazzullo, Saro, Abbott, Abdalla, Bayliss, {Benoit-L{\'e}vy}, Bernstein, Bertin, Brooks, Capasso, Capozzi, Carnero, Carrasco~Kind, Carretero, Chiu, D'Andrea, {daCosta}, Diehl, Doel, Eifler, Evrard, {Fausti-Neto}, Fosalba, Frieman, Gangkofner, Gonzalez, Gruen, Gruendl, Gupta, Gutierrez, Honscheid, {Hlavacek-Larrondo}, James, Kuehn, Kuropatkin, Lahav, March, Marshall, Martini, McDonald, Melchior, Miller, Miquel, Neilsen, Nord, Ogando, Plazas, Reichardt, Romer, Rozo, Rykoff, Sanchez, Santiago, Schubnell, {Sevilla-Noarbe}, Smith, {Soares-Santos}, Sobreira, Stalder, Stanford, Suchyta, Swanson, Tarle, Thomas, Vikram, Walker, \& Zhang}]{hennig_galaxy_2017}
Hennig, C., Mohr, J.~J., Zenteno, A., {et~al.} 2017, \mnras, 467, 4015

\bibitem[{{Hung} {et~al.}(2025){Hung}, {Lemaux}, {Cucciati}, {Forrest}, {Shah}, {Gal}, {Giddings}, {Sikorski}, {Golden-Marx}, {Lubin}, {Hathi}, {Zamorani}, {Shen}, {Bardelli}, {Cassar{\`a}}, {De Lucia}, {Fontanot}, {Garilli}, {Guaita}, {Hirschmann}, {Lee}, {Newman}, {Ramakrishnan}, {Vergani}, {Xie}, \& {Zucca}}]{hung_protocompilation_2025}
{Hung}, D., {Lemaux}, B.~C., {Cucciati}, O., {et~al.} 2025, \apj, 980, 155

\bibitem[{Ilbert {et~al.}(2015)Ilbert, Arnouts, Le~Floc'h, Aussel, Bethermin, Capak, Hsieh, Kajisawa, Karim, Le~F{\`e}vre, Lee, Lilly, McCracken, {Michel-Dansac}, Moutard, Renzini, Salvato, Sanders, Scoville, Sheth, Silverman, Smol{\v c}i{\'c}, Taniguchi, \& Tresse}]{ilbert_evolution_2015}
Ilbert, O., Arnouts, S., Le~Floc'h, E., {et~al.} 2015, \aap, 579, A2

\bibitem[{Ilbert {et~al.}(2006)Ilbert, Arnouts, McCracken, Bolzonella, Bertin, Le~F{\`e}vre, Mellier, Zamorani, Pell{\`o}, Iovino, Tresse, Le~Brun, Bottini, Garilli, Maccagni, Picat, Scaramella, Scodeggio, Vettolani, Zanichelli, Adami, Bardelli, Cappi, Charlot, Ciliegi, Contini, Cucciati, Foucaud, Franzetti, Gavignaud, Guzzo, Marano, Marinoni, Mazure, Meneux, Merighi, Paltani, Pollo, Pozzetti, Radovich, Zucca, Bondi, Bongiorno, Busarello, {de La Torre}, Gregorini, Lamareille, Mathez, Merluzzi, Ripepi, Rizzo, \& Vergani}]{ilbert_accurate_2006}
Ilbert, O., Arnouts, S., McCracken, H.~J., {et~al.} 2006, \aap, 457, 841

\bibitem[{Ilbert {et~al.}(2009)Ilbert, Capak, Salvato, Aussel, McCracken, Sanders, Scoville, Kartaltepe, Arnouts, Le~Floc'h, Mobasher, Taniguchi, Lamareille, Leauthaud, Sasaki, Thompson, Zamojski, Zamorani, Bardelli, Bolzonella, Bongiorno, Brusa, Caputi, Carollo, Contini, Cook, Coppa, Cucciati, {de la Torre}, {de Ravel}, Franzetti, Garilli, Hasinger, Iovino, Kampczyk, Kneib, Knobel, Kovac, Le~Borgne, Le~Brun, Le~F{\`e}vre, Lilly, Looper, Maier, Mainieri, Mellier, Mignoli, Murayama, Pell{\`o}, Peng, {P{\'e}rez-Montero}, Renzini, Ricciardelli, Schiminovich, Scodeggio, Shioya, Silverman, Surace, Tanaka, Tasca, Tresse, Vergani, \& Zucca}]{ilbert_cosmos_2009}
Ilbert, O., Capak, P., Salvato, M., {et~al.} 2009, \apj, 690, 1236

\bibitem[{Ilbert {et~al.}(2013)Ilbert, McCracken, Le~F{\`e}vre, Capak, Dunlop, Karim, Renzini, Caputi, Boissier, Arnouts, Aussel, Comparat, Guo, Hudelot, Kartaltepe, Kneib, Krogager, Le~Floc'h, Lilly, Mellier, {Milvang-Jensen}, Moutard, Onodera, Richard, Salvato, Sanders, Scoville, Silverman, Taniguchi, Tasca, Thomas, Toft, Tresse, Vergani, Wolk, \& Zirm}]{ilbert_mass_2013}
Ilbert, O., McCracken, H.~J., Le~F{\`e}vre, O., {et~al.} 2013, \aap, 556, A55

\bibitem[{Ito {et~al.}(2023)Ito, Tanaka, Valentino, Toft, Brammer, Gould, Ilbert, Kashikawa, Kubo, Liang, McCracken, \& Weaver}]{ito_cosmos2020_2023}
Ito, K., Tanaka, M., Valentino, F., {et~al.} 2023, \apj, 945, L9

\bibitem[{Kartaltepe {et~al.}(2010)Kartaltepe, Sanders, Le~Floc'h, Frayer, Aussel, Arnouts, Ilbert, Salvato, Scoville, Surace, Yan, Capak, Caputi, Carollo, Cassata, Civano, Hasinger, Koekemoer, Le~F{\`e}vre, Lilly, Liu, McCracken, Schinnerer, Smol{\v c}i{\'c}, Taniguchi, Thompson, Trump, Baldassare, \& Fiorenza}]{kartaltepe_multiwavelength_2010}
Kartaltepe, J.~S., Sanders, D.~B., Le~Floc'h, E., {et~al.} 2010, \apj, 721, 98

\bibitem[{Kartaltepe {et~al.}(2015)Kartaltepe, Sanders, Silverman, Kashino, Chu, Zahid, Hasinger, Kewley, Matsuoka, Nagao, Riguccini, Salvato, Schawinski, Taniguchi, Treister, Capak, Daddi, \& Ohta}]{kartaltepe_rest-frame_2015}
Kartaltepe, J.~S., Sanders, D.~B., Silverman, J.~D., {et~al.} 2015, \apj, 806, L35

\bibitem[{Kashino {et~al.}(2019)Kashino, Silverman, Sanders, Kartaltepe, Daddi, Renzini, Rodighiero, Puglisi, Valentino, Juneau, Arimoto, Nagao, Ilbert, Le~F{\`e}vre, \& Koekemoer}]{kashino_fmos-cosmos_2019}
Kashino, D., Silverman, J.~D., Sanders, D., {et~al.} 2019, \apjs, 241, 10

\bibitem[{{Khostovan} {et~al.}(2025){Khostovan}, {Kartaltepe}, {Salvato}, {Ilbert}, {Casey}, {Algera}, {Antwi-Danso}, {Battisti}, {Brinch}, {Brusa}, {Calabro}, {Capak}, {Chartab}, {Cooper}, {Cox}, {Darvish}, {Drakos}, {Faisst}, {George}, {Gozaliasl}, {Harish}, {Hasinger}, {Hatamnia}, {Iovino}, {Jin}, {Kashino}, {Koekemoer}, {Laishram}, {Lee}, {Lertprasertpong}, {Lilly}, {Masters}, {Mobasher}, {Nagao}, {Onodera}, {Peng}, {Sanders}, {Sanders}, {Sattari}, {Scoville}, {Shah}, {Silverman}, {Suzuki}, {Tanaka}, {Toft}, {Trakhtenbrot}, {Trump}, {Vaccari}, {Valentino}, {Vanderhoof}, {Weaver}, {Yun}, \& {Zavala}}]{khostovan_spec_compil_2025}
{Khostovan}, A.~A., {Kartaltepe}, J.~S., {Salvato}, M., {et~al.} 2025, arXiv e-prints, arXiv:2503.00120

\bibitem[{{Kiyota} {et~al.}(2025){Kiyota}, {Ando}, {Tanaka}, {Finoguenov}, {Ali}, {Coupon}, {Desprez}, {Gwyn}, {Sawicki}, \& {Shimakawa}}]{kiyota_cluster_quiescent_2025}
{Kiyota}, T., {Ando}, M., {Tanaka}, M., {et~al.} 2025, \apj, 980, 104

\bibitem[{Knobel {et~al.}(2012)Knobel, Lilly, Iovino, Kova{\v c}, Bschorr, Presotto, Oesch, Kampczyk, Carollo, Contini, Kneib, Le~Fevre, Mainieri, Renzini, Scodeggio, Zamorani, Bardelli, Bolzonella, Bongiorno, Caputi, Cucciati, {de la Torre}, {de Ravel}, Franzetti, Garilli, Lamareille, Le~Borgne, Le~Brun, Maier, Mignoli, Pello, Peng, Perez~Montero, Silverman, Tanaka, Tasca, Tresse, Vergani, Zucca, Barnes, Bordoloi, Cappi, Cimatti, Coppa, Koekemoer, {L{\'o}pez-Sanjuan}, McCracken, Moresco, Nair, Pozzetti, \& Welikala}]{knobel_zcosmos_2012}
Knobel, C., Lilly, S.~J., Iovino, A., {et~al.} 2012, \apj, 753, 121

\bibitem[{Koekemoer {et~al.}(2007)Koekemoer, Aussel, Calzetti, Capak, Giavalisco, Kneib, Leauthaud, Le~F{\`e}vre, McCracken, Massey, Mobasher, Rhodes, Scoville, \& Shopbell}]{koekemoer_cosmos_2007}
Koekemoer, A.~M., Aussel, H., Calzetti, D., {et~al.} 2007, \apjs, 172, 196

\bibitem[{Kotulla {et~al.}(2009)Kotulla, Fritze, Weilbacher, \& Anders}]{kotulla_galev_2009}
Kotulla, R., Fritze, U., Weilbacher, P., \& Anders, P. 2009, \mnras, 396, 462

\bibitem[{Koyama {et~al.}(2021)Koyama, Polletta, Tanaka, Kodama, Dole, Soucail, Frye, Lehnert, \& Scodeggio}]{koyama_planck-selected_2021}
Koyama, Y., Polletta, M. d.~C., Tanaka, I., {et~al.} 2021, \mnras, 503, L1

\bibitem[{Kukstas {et~al.}(2023)Kukstas, Balogh, McCarthy, Bah{\'e}, De~Lucia, Jablonka, Vulcani, Baxter, Biviano, Cerulo, Chan, Cooper, Demarco, Finoguenov, Font, Lidman, Marchioni, McGee, Muzzin, Nantais, Old, {Pintos-Castro}, Poggianti, Reeves, Rudnick, Sarron, {van der Burg}, Webb, Wilson, Yee, \& Zaritsky}]{kukstas_gogreen_2023}
Kukstas, E., Balogh, M.~L., McCarthy, I.~G., {et~al.} 2023, \mnras, 518, 4782

\bibitem[{Laigle {et~al.}(2016)Laigle, McCracken, Ilbert, Hsieh, Davidzon, Capak, Hasinger, Silverman, Pichon, Coupon, Aussel, Le~Borgne, Caputi, Cassata, Chang, Civano, Dunlop, Fynbo, Kartaltepe, Koekemoer, Le~F{\`e}vre, Le~Floc'h, Leauthaud, Lilly, Lin, Marchesi, {Milvang-Jensen}, Salvato, Sanders, Scoville, Smolcic, Stockmann, Taniguchi, Tasca, Toft, Vaccari, \& Zabl}]{laigle_cosmos2015_2016}
Laigle, C., McCracken, H.~J., Ilbert, O., {et~al.} 2016, \apjs, 224, 24

\bibitem[{Laigle {et~al.}(2018)Laigle, Pichon, Arnouts, McCracken, Dubois, Devriendt, Slyz, Le~Borgne, {Benoit-L{\'e}vy}, Hwang, Ilbert, Kraljic, Malavasi, Park, \& Vibert}]{laigle_cosmos2015_2018}
Laigle, C., Pichon, C., Arnouts, S., {et~al.} 2018, \mnras, 474, 5437

\bibitem[{Laureijs {et~al.}(2011)Laureijs, Amiaux, Arduini, Augu{\`e}res, Brinchmann, Cole, Cropper, Dabin, Duvet, Ealet, Garilli, Gondoin, Guzzo, Hoar, Hoekstra, Holmes, Kitching, Maciaszek, Mellier, Pasian, Percival, Rhodes, Saavedra~Criado, Sauvage, Scaramella, Valenziano, Warren, Bender, Castander, Cimatti, Le~F{\`e}vre, {Kurki-Suonio}, Levi, Lilje, Meylan, Nichol, Pedersen, Popa, Rebolo~Lopez, Rix, Rottgering, Zeilinger, Grupp, Hudelot, Massey, Meneghetti, Miller, Paltani, {Paulin-Henriksson}, Pires, Saxton, Schrabback, Seidel, Walsh, Aghanim, Amendola, Bartlett, Baccigalupi, Beaulieu, Benabed, Cuby, Elbaz, Fosalba, Gavazzi, Helmi, Hook, Irwin, Kneib, Kunz, Mannucci, Moscardini, Tao, Teyssier, Weller, Zamorani, Zapatero~Osorio, Boulade, Foumond, Di~Giorgio, Guttridge, James, Kemp, Martignac, Spencer, Walton, Bl{\"u}mchen, Bonoli, Bortoletto, Cerna, Corcione, Fabron, Jahnke, Ligori, Madrid, Martin, Morgante, Pamplona, Prieto, Riva, Toledo, Trifoglio, Zerbi, Abdalla, Douspis, Grenet, Borgani, Bouwens,
  Courbin, Delouis, Dubath, Fontana, Frailis, Grazian, Koppenh{\"o}fer, Mansutti, Melchior, Mignoli, Mohr, Neissner, Noddle, Poncet, Scodeggio, Serrano, Shane, Starck, Surace, Taylor, {Verdoes-Kleijn}, Vuerli, Williams, Zacchei, Altieri, Escudero~Sanz, Kohley, Oosterbroek, Astier, Bacon, Bardelli, Baugh, Bellagamba, Benoist, Bianchi, Biviano, Branchini, Carbone, Cardone, Clements, Colombi, Conselice, Cresci, Deacon, Dunlop, Fedeli, Fontanot, Franzetti, Giocoli, {Garcia-Bellido}, Gow, Heavens, Hewett, Heymans, Holland, Huang, Ilbert, Joachimi, Jennins, Kerins, Kiessling, Kirk, Kotak, Krause, Lahav, {van Leeuwen}, Lesgourgues, Lombardi, Magliocchetti, Maguire, Majerotto, Maoli, Marulli, Maurogordato, McCracken, McLure, Melchiorri, Merson, Moresco, Nonino, Norberg, Peacock, Pello, Penny, Pettorino, Di~Porto, Pozzetti, Quercellini, Radovich, Rassat, Roche, Ronayette, \& Rossetti}]{laureijs_euclid_2011}
Laureijs, R., Amiaux, J., Arduini, S., {et~al.} 2011, Euclid {{Definition Study Report}}

\bibitem[{Lee {et~al.}(2014)Lee, Hennawi, Stark, Prochaska, White, Schlegel, Eilers, {Arinyo-i-Prats}, Suzuki, Croft, Caputi, Cassata, Ilbert, Garilli, Koekemoer, Le~Brun, Le~F{\`e}vre, Maccagni, Nugent, Taniguchi, Tasca, Tresse, Zamorani, \& Zucca}]{lee_ly_2014}
Lee, K.-G., Hennawi, J.~F., Stark, C., {et~al.} 2014, \apj, 795, L12

\bibitem[{Lee {et~al.}(2016)Lee, Hennawi, White, Prochaska, {Font-Ribera}, Schlegel, Rich, Suzuki, Stark, Le~F{\`e}vre, Nugent, Salvato, \& Zamorani}]{lee_shadow_2016}
Lee, K.-G., Hennawi, J.~F., White, M., {et~al.} 2016, \apj, 817, 160

\bibitem[{Lietzen {et~al.}(2012)Lietzen, Tempel, Hein{\"a}m{\"a}ki, Nurmi, Einasto, \& Saar}]{lietzen_environments_2012}
Lietzen, H., Tempel, E., Hein{\"a}m{\"a}ki, P., {et~al.} 2012, \aap, 545, A104

\bibitem[{Lilly {et~al.}(2009)Lilly, Le~Brun, Maier, Mainieri, Mignoli, Scodeggio, Zamorani, Carollo, Contini, Kneib, Le~F{\`e}vre, Renzini, Bardelli, Bolzonella, Bongiorno, Caputi, Coppa, Cucciati, {de la Torre}, {de Ravel}, Franzetti, Garilli, Iovino, Kampczyk, Kovac, Knobel, Lamareille, Le~Borgne, Pello, Peng, {P{\'e}rez-Montero}, Ricciardelli, Silverman, Tanaka, Tasca, Tresse, Vergani, Zucca, Ilbert, Salvato, Oesch, Abbas, Bottini, Capak, Cappi, Cassata, Cimatti, Elvis, Fumana, Guzzo, Hasinger, Koekemoer, Leauthaud, Maccagni, Marinoni, McCracken, Memeo, Meneux, Porciani, Pozzetti, Sanders, Scaramella, Scarlata, Scoville, Shopbell, \& Taniguchi}]{lilly_zcosmos_2009}
Lilly, S.~J., Le~Brun, V., Maier, C., {et~al.} 2009, \apjs, 184, 218

\bibitem[{Lilly {et~al.}(2007)Lilly, Le~F{\`e}vre, Renzini, Zamorani, Scodeggio, Contini, Carollo, Hasinger, Kneib, Iovino, Le~Brun, Maier, Mainieri, Mignoli, Silverman, Tasca, Bolzonella, Bongiorno, Bottini, Capak, Caputi, Cimatti, Cucciati, Daddi, Feldmann, Franzetti, Garilli, Guzzo, Ilbert, Kampczyk, Kovac, Lamareille, Leauthaud, Le~Borgne, McCracken, Marinoni, Pello, Ricciardelli, Scarlata, Vergani, Sanders, Schinnerer, Scoville, Taniguchi, Arnouts, Aussel, Bardelli, Brusa, Cappi, Ciliegi, Finoguenov, Foucaud, Franceschini, Halliday, Impey, Knobel, Koekemoer, Kurk, Maccagni, Maddox, Marano, Marconi, Meneux, Mobasher, Moreau, Peacock, Porciani, Pozzetti, Scaramella, Schiminovich, Shopbell, Smail, Thompson, Tresse, Vettolani, Zanichelli, \& Zucca}]{lilly_zcosmos_2007}
Lilly, S.~J., Le~F{\`e}vre, O., Renzini, A., {et~al.} 2007, \apjs, 172, 70

\bibitem[{Lovisari {et~al.}(2021)Lovisari, Ettori, Gaspari, \& Giles}]{lovisari_scaling_2021}
Lovisari, L., Ettori, S., Gaspari, M., \& Giles, P.~A. 2021, Universe, 7, 139

\bibitem[{Mandelbaum {et~al.}(2006)Mandelbaum, Seljak, Cool, Blanton, Hirata, \& Brinkmann}]{mandelbaum_density_2006}
Mandelbaum, R., Seljak, U., Cool, R.~J., {et~al.} 2006, \mnras, 372, 758

\bibitem[{Maturi {et~al.}(2019)Maturi, Bellagamba, Radovich, Roncarelli, Sereno, Moscardini, Bardelli, \& Puddu}]{maturi_amico_2019}
Maturi, M., Bellagamba, F., Radovich, M., {et~al.} 2019, \mnras, 485, 498

\bibitem[{Maturi {et~al.}(2023)Maturi, Finoguenov, Lopes, Gonz{\'a}lez~Delgado, Dupke, Cypriano, Carrasco, Diego, {Penna-Lima}, Doubrawa, V{\'i}lchez, Moscardini, Marra, Bonoli, {Rodr{\'i}guez-Mart{\'i}n}, Zitrin, M{\'a}rquez, {Hern{\'a}n-Caballero}, {Jim{\'e}nez-Teja}, Abramo, Alcaniz, Benitez, Carneiro, Cenarro, {Crist{\'o}bal-Hornillos}, Ederoclite, {L{\'o}pez-Sanjuan}, {Mar{\'i}n-Franch}, {Mendes de Oliveira}, Moles, Sodr{\'e}, Taylor, Varela, V{\'a}zquez~Rami{\'o}, \& {Fern{\'a}ndez-Ontiveros}}]{maturi_minijpas_2023}
Maturi, M., Finoguenov, A., Lopes, P. A.~A., {et~al.} 2023, \aap, 678, A145

\bibitem[{Maturi {et~al.}(2005)Maturi, Meneghetti, Bartelmann, Dolag, \& Moscardini}]{maturi_optimal_2005}
Maturi, M., Meneghetti, M., Bartelmann, M., Dolag, K., \& Moscardini, L. 2005, \aap, 442, 851

\bibitem[{McCarthy {et~al.}(2010)McCarthy, Schaye, Ponman, Bower, Booth, Dalla~Vecchia, Crain, Springel, Theuns, \& Wiersma}]{mccarthy_case_2010}
McCarthy, I.~G., Schaye, J., Ponman, T.~J., {et~al.} 2010, \mnras, 406, 822

\bibitem[{McConachie {et~al.}(2022)McConachie, Wilson, Forrest, Marsan, Muzzin, Cooper, Annunziatella, Marchesini, Chan, Gomez, Abdullah, Saracco, \& Nantais}]{mcconachie_spectroscopic_2022}
McConachie, I., Wilson, G., Forrest, B., {et~al.} 2022, \apj, 926, 37

\bibitem[{McConachie {et~al.}(2025)McConachie, Wilson, Forrest, Marsan, Muzzin, Cooper, Annunziatella, Marchesini, Gomez, Chang, Urbano~Stawinski, McDonald, Webb, Noble, Lemaux, Shah, Staab, Lubin, \& Gal}]{mcconachie_magaz3ne_2025}
McConachie, I., Wilson, G., Forrest, B., {et~al.} 2025, \apj, 978, 17

\bibitem[{McCracken {et~al.}(2012)McCracken, {Milvang-Jensen}, Dunlop, Franx, Fynbo, Le~F{\`e}vre, Holt, Caputi, Goranova, Buitrago, Emerson, Freudling, Hudelot, {L{\'o}pez-Sanjuan}, Magnard, Mellier, M{\o}ller, Nilsson, Sutherland, Tasca, \& Zabl}]{mccracken_ultravista_2012}
McCracken, H.~J., {Milvang-Jensen}, B., Dunlop, J., {et~al.} 2012, \aap, 544, A156

\bibitem[{McGee {et~al.}(2009)McGee, Balogh, Bower, Font, \& McCarthy}]{mcgee_accretion_2009}
McGee, S.~L., Balogh, M.~L., Bower, R.~G., Font, A.~S., \& McCarthy, I.~G. 2009, \mnras, 400, 937

\bibitem[{McNab {et~al.}(2021)McNab, Balogh, {van der Burg}, Forestell, Webb, Vulcani, Rudnick, Muzzin, Cooper, McGee, Biviano, Cerulo, Chan, De~Lucia, Demarco, Finoguenov, Forrest, Golledge, Jablonka, Lidman, Nantais, Old, {Pintos-Castro}, Poggianti, Reeves, Wilson, Yee, \& Zaritsky}]{mcnab_gogreen_2021}
McNab, K., Balogh, M.~L., {van der Burg}, R. F.~J., {et~al.} 2021, \mnras, 508, 157

\bibitem[{Mellier {et~al.}(2018)Mellier, Racca, \& Laureijs}]{mellier_unveiling_2018}
Mellier, Y., Racca, G., \& Laureijs, R. 2018, 42, E1.16

\bibitem[{Moneti {et~al.}(2023)Moneti, McCracken, Hudelot, Rouberol, Herent, Mellier, Dunlop, Le~Fevre, Franx, Fynbo, Bowler, Caputi, Kauffmann, {Milvang-Jensen}, {Gonzalez-Fernandez}, {Gonzalez-Solares}, Irwin, Lewis, Blake, Cross, Read, \& Sutorius}]{moneti_vizier_2023}
Moneti, A., McCracken, H.~J., Hudelot, W., {et~al.} 2023, VizieR Online Data Cat., 2373, II/373

\bibitem[{{Morishita} {et~al.}(2024){Morishita}, {Liu}, {Stiavelli}, {Treu}, {Trenti}, {Chartab}, {Roberts-Borsani}, {Vulcani}, {Bergamini}, {Castellano}, \& {Grillo}}]{morishita_earlyoverdensity_2024}
{Morishita}, T., {Liu}, Z., {Stiavelli}, M., {et~al.} 2024, arXiv e-prints, arXiv:2408.10980

\bibitem[{{Morishita} {et~al.}(2023){Morishita}, {Roberts-Borsani}, {Treu}, {Brammer}, {Mason}, {Trenti}, {Vulcani}, {Wang}, {Acebron}, {Bah{\'e}}, {Bergamini}, {Boyett}, {Bradac}, {Calabr{\`o}}, {Castellano}, {Chen}, {De Lucia}, {Filippenko}, {Fontana}, {Glazebrook}, {Grillo}, {Henry}, {Jones}, {Kelly}, {Koekemoer}, {Leethochawalit}, {Lu}, {Marchesini}, {Mascia}, {Mercurio}, {Merlin}, {Metha}, {Nanayakkara}, {Nonino}, {Paris}, {Pentericci}, {Rosati}, {Santini}, {Strait}, {Vanzella}, {Windhorst}, \& {Xie}}]{morishita_protocluster_jwst_2023}
{Morishita}, T., {Roberts-Borsani}, G., {Treu}, T., {et~al.} 2023, \apjl, 947, L24

\bibitem[{Navarro {et~al.}(1997)Navarro, Frenk, \& White}]{navarro_universal_1997}
Navarro, J.~F., Frenk, C.~S., \& White, S. D.~M. 1997, \apj, 490, 493

\bibitem[{{Newman} {et~al.}(2022){Newman}, {Rudie}, {Blanc}, {Qezlou}, {Bird}, {Kelson}, {P{\'e}rez}, {Congiu}, {Lemaux}, {Dressler}, \& {Mulchaey}}]{newman_protocl_2022}
{Newman}, A.~B., {Rudie}, G.~C., {Blanc}, G.~A., {et~al.} 2022, \nat, 606, 475

\bibitem[{Paul {et~al.}(2017)Paul, John, Gupta, \& Kumar}]{paul_understanding_2017}
Paul, S., John, R.~S., Gupta, P., \& Kumar, H. 2017, \mnras, 471, 2

\bibitem[{Polletta {et~al.}(2021)Polletta, Soucail, Dole, Lehnert, Pointecouteau, Vietri, Scodeggio, Montier, Koyama, Lagache, Frye, Cusano, \& Fumana}]{polletta_spectroscopic_2021}
Polletta, M., Soucail, G., Dole, H., {et~al.} 2021, \aap, 654, A121

\bibitem[{Puddu {et~al.}(2021)Puddu, Radovich, Sereno, Bardelli, Maturi, Moscardini, Bellagamba, Giocoli, Marulli, \& Roncarelli}]{puddu_amico_2021}
Puddu, E., Radovich, M., Sereno, M., {et~al.} 2021, \aap, 645, A9

\bibitem[{Ragagnin {et~al.}(2021)Ragagnin, Saro, Singh, \& Dolag}]{ragagnin_cosmology_2021}
Ragagnin, A., Saro, A., Singh, P., \& Dolag, K. 2021, \mnras, 500, 5056

\bibitem[{Reeves {et~al.}(2021)Reeves, Balogh, {van der Burg}, Finoguenov, Kukstas, McCarthy, Webb, Muzzin, McGee, Rudnick, Biviano, Cerulo, Chan, Cooper, Demarco, Jablonka, De~Lucia, Vulcani, Wilson, Yee, \& Zaritsky}]{reeves_gogreen_2021}
Reeves, A. M.~M., Balogh, M.~L., {van der Burg}, R. F.~J., {et~al.} 2021, \mnras, 506, 3364

\bibitem[{Rieke {et~al.}(2023)Rieke, Kelly, Misselt, Stansberry, Boyer, Beatty, Egami, Florian, Greene, Hainline, Leisenring, Roellig, Schlawin, Sun, Tinnin, Williams, Willmer, Wilson, Clark, Rohrbach, Brooks, Canipe, Correnti, DiFelice, Gennaro, Girard, Hartig, Hilbert, Koekemoer, Nikolov, Pirzkal, Rest, Robberto, Sunnquist, Telfer, Wu, Ferry, Lewis, Baum, Beichman, Doyon, Dressler, Eisenstein, Ferrarese, Hodapp, Horner, Jaffe, Johnstone, Krist, Martin, McCarthy, Meyer, Rieke, Trauger, \& Young}]{rieke_performance_2023}
Rieke, M.~J., Kelly, D.~M., Misselt, K., {et~al.} 2023, \pasp, 135, 028001

\bibitem[{Salerno {et~al.}(2019)Salerno, Mart{\'i}nez, \& Muriel}]{salerno_filaments_2019}
Salerno, J.~M., Mart{\'i}nez, H.~J., \& Muriel, H. 2019, \mnras, 484, 2

\bibitem[{Sarron \& Conselice(2021)}]{sarron_detectifz_2021}
Sarron, F. \& Conselice, C.~J. 2021, \mnras, 506, 2136

\bibitem[{Sawicki {et~al.}(2019)Sawicki, Arnouts, Huang, Coupon, Golob, Gwyn, Foucaud, Moutard, Iwata, Liu, Chen, Desprez, Harikane, Ono, Strauss, Tanaka, Thibert, Balogh, Bundy, Chapman, Gunn, Hsieh, Ilbert, Jing, LeF{\`e}vre, Li, Matsuda, Miyazaki, Nagao, Nishizawa, Ouchi, Shimasaku, Silverman, {de la Torre}, Tresse, Wang, Willott, Yamada, Yang, \& Yee}]{sawicki_cfht_2019}
Sawicki, M., Arnouts, S., Huang, J., {et~al.} 2019, \mnras, 489, 5202

\bibitem[{Schechter(1976)}]{schechter_analytic_1976}
Schechter, P. 1976, \apj, 203, 297

\bibitem[{Scoville {et~al.}(2013)Scoville, Arnouts, Aussel, Benson, Bongiorno, Bundy, Calvo, Capak, Carollo, Civano, Dunlop, Elvis, Faisst, Finoguenov, Fu, Giavalisco, Guo, Ilbert, Iovino, Kajisawa, Kartaltepe, Leauthaud, Le~F{\`e}vre, LeFloch, Lilly, Liu, Manohar, Massey, Masters, McCracken, Mobasher, Peng, Renzini, Rhodes, Salvato, Sanders, Sarvestani, Scarlata, Schinnerer, Sheth, Shopbell, Smol{\v c}i{\'c}, Taniguchi, Taylor, White, \& Yan}]{scoville_evolution_2013}
Scoville, N., Arnouts, S., Aussel, H., {et~al.} 2013, \apjs, 206, 3

\bibitem[{Scoville {et~al.}(2007)Scoville, Aussel, Brusa, Capak, Carollo, Elvis, Giavalisco, Guzzo, Hasinger, Impey, Kneib, LeFevre, Lilly, Mobasher, Renzini, Rich, Sanders, Schinnerer, Schminovich, Shopbell, Taniguchi, \& Tyson}]{scoville_cosmic_2007}
Scoville, N., Aussel, H., Brusa, M., {et~al.} 2007, \apjs, 172, 1

\bibitem[{S{\'e}rsic(1963)}]{sersic_influence_1963}
S{\'e}rsic, J.~L. 1963, Boletin Asoc. Argent. Astron. Plata Argent., 6, 41

\bibitem[{Shimakawa {et~al.}(2018)Shimakawa, Koyama, R{\"o}ttgering, Kodama, Hayashi, Hatch, Dannerbauer, Tanaka, Tadaki, Suzuki, Fukagawa, Cai, \& Kurk}]{shimakawa_mahalo_2018}
Shimakawa, R., Koyama, Y., R{\"o}ttgering, H. J.~A., {et~al.} 2018, \mnras, 481, 5630

\bibitem[{{Shuntov} {et~al.}(2025){Shuntov}, {Ilbert}, {Toft}, {Arango-Toro}, {Akins}, {Casey}, {Franco}, {Harish}, {Kartaltepe}, {Koekemoer}, {McCracken}, {Paquereau}, {Laigle}, {Bethermin}, {Dubois}, {Drakos}, {Faisst}, {Gozaliasl}, {Gillman}, {Hayward}, {Hirschmann}, {Huertas-Company}, {Jespersen}, {Jin}, {Kokorev}, {Lambrides}, {Le Borgne}, {Liu}, {Magdis}, {Massey}, {McPartland}, {Mercier}, {McCleary}, {McKinney}, {Oesch}, {Renzini}, {Rhodes}, {Rich}, {Robertson}, {Sanders}, {Trebitsch}, {Tresse}, {Valentino}, {Vijayan}, {Weaver}, {Weibel}, {Wilkins}, \& {Yang}}]{shuntov_cosmos-web_2024}
{Shuntov}, M., {Ilbert}, O., {Toft}, S., {et~al.} 2025, \aap, 695, A20

\bibitem[{Sillassen {et~al.}(2024)Sillassen, Jin, Magdis, Daddi, Wang, Lu, Sun, Arumugam, Liu, Brinch, D'Eugenio, Gobat, {G{\'o}mez-Guijarro}, Rich, Schinnerer, Strazzullo, Tan, Valentino, Wang, Xiao, Zhou, {Bl{\'a}nquez-Ses{\'e}}, Cai, Chen, Ciesla, Dai, Delvecchio, Elbaz, Finoguenov, Gao, Gu, Hale, Hao, Huang, Jarvis, Kalita, Ke, Le~Bail, Magnelli, Shi, Vaccari, Whittam, Yang, \& Zhang}]{sillassen_noema_2024}
Sillassen, N.~B., Jin, S., Magdis, G.~E., {et~al.} 2024, \aap, 690, A55

\bibitem[{Silverman {et~al.}(2015)Silverman, Kashino, Sanders, Kartaltepe, Arimoto, Renzini, Rodighiero, Daddi, Zahid, Nagao, Kewley, Lilly, Sugiyama, Baronchelli, Capak, Carollo, Chu, Hasinger, Ilbert, Juneau, Kajisawa, Koekemoer, Kovac, Le~F{\`e}vre, Masters, McCracken, Onodera, Schulze, Scoville, Strazzullo, \& Taniguchi}]{silverman_fmos-cosmos_2015}
Silverman, J.~D., Kashino, D., Sanders, D., {et~al.} 2015, \apjs, 220, 12

\bibitem[{Smol{\v c}i{\'c} {et~al.}(2017)Smol{\v c}i{\'c}, Novak, Bondi, Ciliegi, Mooley, Schinnerer, Zamorani, Navarrete, Bourke, Karim, Vardoulaki, Leslie, Delhaize, Carilli, Myers, Baran, Delvecchio, Miettinen, Banfield, Balokovi{\'c}, Bertoldi, Capak, Frail, Hallinan, Hao, Herrera~Ruiz, Horesh, Ilbert, Intema, Jeli{\'c}, Kl{\"o}ckner, Krpan, Kulkarni, McCracken, Laigle, Middleberg, Murphy, Sargent, Scoville, \& Sheth}]{smolcic_vla-cosmos_2017}
Smol{\v c}i{\'c}, V., Novak, M., Bondi, M., {et~al.} 2017, \aap, 602, A1

\bibitem[{Spitler {et~al.}(2012)Spitler, Labb{\'e}, Glazebrook, Persson, Monson, Papovich, Tran, Poole, Quadri, {van Dokkum}, Kelson, Kacprzak, McCarthy, Murphy, Straatman, \& Tilvi}]{spitler_first_2012}
Spitler, L.~R., Labb{\'e}, I., Glazebrook, K., {et~al.} 2012, \apj, 748, L21

\bibitem[{{Taamoli} {et~al.}(2024){Taamoli}, {Nezhad}, {Mobasher}, {Manesh}, {Chartab}, {Weaver}, {Capak}, {Casey}, {Gozaliasl}, {Heintz}, {Ilbert}, {Kartaltepe}, {McCracken}, {Sanders}, {Scoville}, {Toft}, \& {Watson}}]{taamoli_cosmos2020_2024}
{Taamoli}, S., {Nezhad}, N., {Mobasher}, B., {et~al.} 2024, \apj, 977, 263

\bibitem[{Tanaka {et~al.}(2012)Tanaka, Finoguenov, Lilly, Bolzonella, Carollo, Contini, Iovino, Kneib, Lamareille, Le~Fevre, Mainieri, Presotto, Renzini, Scodeggio, Silverman, Zamorani, Bardelli, Bongiorno, Caputi, Cucciati, {de la Torre}, {de Ravel}, Franzetti, Garilli, Kampczyk, Knobel, Kova{\v c}, Le~Borgne, Le~Brun, {L{\'o}pez-Sanjuan}, Maier, Mignoli, Pello, Peng, {Perez-Montero}, Tasca, Tresse, Vergani, Zucca, Barnes, Bordoloi, Cappi, Cimatti, Coppa, Koekemoer, McCracken, Moresco, Nair, Oesch, Pozzetti, \& Welikala}]{tanaka_x-ray_2012}
Tanaka, M., Finoguenov, A., Lilly, S.~J., {et~al.} 2012, \pasj, 64, 22

\bibitem[{Taniguchi {et~al.}(2015)Taniguchi, Kajisawa, Kobayashi, Shioya, Nagao, Capak, Aussel, Ichikawa, Murayama, Scoville, Ilbert, Salvato, Sanders, Mobasher, Miyazaki, Komiyama, Le~F{\`e}vre, Tasca, Lilly, Carollo, Renzini, Rich, Schinnerer, Kaifu, Karoji, Arimoto, Okamura, Ohta, Shimasaku, \& Hayashino}]{taniguchi_subaru_2015}
Taniguchi, Y., Kajisawa, M., Kobayashi, M. A.~R., {et~al.} 2015, \pasj, 67, 104

\bibitem[{Toni {et~al.}(2024)Toni, Maturi, Finoguenov, Moscardini, \& Castignani}]{toni_amico-cosmos_2024}
Toni, G., Maturi, M., Finoguenov, A., Moscardini, L., \& Castignani, G. 2024, \aap, 687, A56

\bibitem[{Tully(1987)}]{tully_nearby_1987}
Tully, R.~B. 1987, \apj, 321, 280

\bibitem[{Vulcani {et~al.}(2018)Vulcani, Poggianti, Jaff{\'e}, Moretti, Fritz, Gullieuszik, Bettoni, Fasano, Tonnesen, \& McGee}]{vulcani_gasp_2018}
Vulcani, B., Poggianti, B.~M., Jaff{\'e}, Y.~L., {et~al.} 2018, \mnras, 480, 3152

\bibitem[{Wang {et~al.}(2016)Wang, Elbaz, Daddi, Finoguenov, Liu, Schreiber, Mart{\'i}n, Strazzullo, Valentino, {van der Burg}, Zanella, Ciesla, Gobat, Le~Brun, Pannella, Sargent, Shu, Tan, Cappelluti, \& Li}]{wang_discovery_2016}
Wang, T., Elbaz, D., Daddi, E., {et~al.} 2016, \apj, 828, 56

\bibitem[{Weaver {et~al.}(2022)Weaver, Kauffmann, Ilbert, McCracken, Moneti, Toft, Brammer, Shuntov, Davidzon, Hsieh, Laigle, Anastasiou, Jespersen, Vinther, Capak, Casey, McPartland, {Milvang-Jensen}, Mobasher, Sanders, Zalesky, Arnouts, Aussel, Dunlop, Faisst, Franx, Furtak, Fynbo, Gould, Greve, Gwyn, Kartaltepe, Kashino, Koekemoer, Kokorev, Le~F{\`e}vre, Lilly, Masters, Magdis, Mehta, Peng, Riechers, Salvato, Sawicki, Scarlata, Scoville, Shirley, Silverman, Sneppen, Smol{\u c}i{\'c}, Steinhardt, Stern, Tanaka, Taniguchi, Teplitz, Vaccari, Wang, \& Zamorani}]{weaver_cosmos2020_2022}
Weaver, J.~R., Kauffmann, O.~B., Ilbert, O., {et~al.} 2022, \apjs, 258, 11

\bibitem[{Wilman {et~al.}(2005)Wilman, Balogh, Bower, Mulchaey, Oemler, Carlberg, Morris, \& Whitaker}]{wilman_galaxy_2005}
Wilman, D.~J., Balogh, M.~L., Bower, R.~G., {et~al.} 2005, \mnras, 358, 71

\bibitem[{Wright {et~al.}(2022)Wright, Sabatke, \& Telfer}]{wright_james_2022}
Wright, R.~H., Sabatke, D., \& Telfer, R. 2022, 12180, 121803P

\bibitem[{Yuan {et~al.}(2014)Yuan, Nanayakkara, Kacprzak, Tran, Glazebrook, Kewley, Spitler, Poole, Labb{\'e}, Straatman, \& Tomczak}]{yuan_keckmosfire_2014}
Yuan, T., Nanayakkara, T., Kacprzak, G.~G., {et~al.} 2014, \apj, 795, L20

\bibitem[{Zhang {et~al.}(2020)Zhang, {Ramos-Ceja}, Pacaud, \& Reiprich}]{zhang_high-redshift_2020}
Zhang, C., {Ramos-Ceja}, M.~E., Pacaud, F., \& Reiprich, T.~H. 2020, \aap, 642, A17

\end{thebibliography}

\begin{appendix}
\section{Structure of the catalog}\label{appendixA}
We summarize the structure of the group catalog with the description of each column in Table \ref{descrcols}.
\begin{table*}[htbp]\label{descrcols}
    \centering
    \begin{tabular}{l l l l l}
        \hline
        \hline
        \noalign{\smallskip}
        Catalog & Column Name & Description & Range \\
        \noalign{\smallskip}
        \hline
        \noalign{\smallskip}
        \textbf{Group catalog} & \texttt{ID} (1) & Group identification number & 1--1816 \\
        & \texttt{RA} (2)  & Right Ascension (R.A.) in degrees & 149.66--150.56 \\
        & \texttt{DEC} (3)  & Declination (Dec.) in degrees & 1.73--2.67 \\
        & \texttt{Z} (4) & AMICO group redshift & 0.08--3.70 \\
        & \texttt{SN} (5) & $S/N$ (incl. background + cluster) & 1.28--6.65 \\
        & \texttt{SN\_NOCL} (6) & $S/N_{nocl}$ ($S/N$ incl. only background) & 6.00--43.11 \\
        & \texttt{AMP} (7) & Signal amplitude & 0.12--2.51 \\
        & \texttt{MSKFRC} (8) & Masked fraction & 0.03--0.79 \\
        & \texttt{LAMBDA} (9) & (Apparent) richness & 14--456 \\
        & \texttt{LAMBDA\_STAR} (10) & (Intrinsic) richness ($m<m_*+1.5$, $r<R_{200}$) & 2.0--77.6 \\
        & \texttt{DETECTION\_FLAG} (11) & Detection flag (see Sect. \ref{qualityflags}) & 0--120 \\
        & \texttt{Z\_SPEC} (12) & Mean spectroscopic redshift (if available) & 0.08--3.80 \\
        & \texttt{N\_SPEC} (13) & Number of members with spectroscopic redshift & 0--103 \\
        & \texttt{ZPHYS\_SIGM} (14) & 1$\sigma$ lower uncertainty on redshift & 0.004-0.281 \\
        & \texttt{ZPHYS\_SIGP} (15) & 1$\sigma$ upper uncertainty on redshift & 0.000-0.192 \\
        \noalign{\smallskip}
        \hline
        \noalign{\smallskip}
        \textbf{Member catalog} &\texttt{GALID (1)} & COSMOS-Web galaxy identification number & 0--783999 \\
        & \texttt{FIELD\_PROB} (2) & Probability to belong to the field & 0--0.995 \\
        & \texttt{ID} (3) & Group identification number & 1--1816 \\
        & \texttt{ASSOC\_PROB} (4) & Probability to belong to the  group indicated in "\texttt{ID}" & 0.005--1 \\
        \noalign{\smallskip}
        \hline
        \hline 
        \noalign{\smallskip}
    \end{tabular}
    \caption{Descriptions of columns for the group and member catalog.}
    \label{cols}
\end{table*}

\section{AMICO run with spectroscopic redshifts}\label{appendixB}
The availability of almost 100 spectroscopic surveys covering the COSMOS field and collected in the compilation by \cite{khostovan_spec_compil_2025} makes this an ideal chance to test cluster and group detection with AMICO, including the information coming from spectroscopic redshifts.

In Sect. \ref{bias}, we performed a simple a-posteriori association of spectroscopic counterparts to the member galaxies retrieved with the AMICO detection described in Sect. \ref{amico}, which is entirely based on photometric redshifts. In this Appendix, we describe instead a group search done with AMICO by using not only photometric redshifts but also spectroscopic redshifts when available. First of all, we associated spectroscopic redshifts with the full cleaned input galaxy catalog. This yielded a galaxy sample with spectroscopic redshift available for almost 20,000 galaxies, which is 5\% of the initial input catalog. Besides these, we kept all galaxies with the photo-$z$, only creating a hybrid input catalog. We built this hybrid catalog by selecting galaxies with spec-$z$ and by shifting their redshift to the spectroscopic values. Then, we added a 1$\sigma$ error, compatible with the size of the chosen AMICO redshift resolution, namely $\Delta z=0.01(1+z)$. We then run AMICO detection on the hybrid catalog with the same parameters described in Sect. \ref{amico}.
This run resulted in a catalog of 1559 candidate clusters and groups in the range $0.03 \leq z \leq 3.7$ and with $\lambda_\star > 2$. Among these, a total of 595 groups are detected with $S/N_{nocl}>10.0$.
We compared the results of this run with the one based only on photo-$z$ by matching the two catalogs within d$z$ = 0.05(1+$z$) and d$r$ = 1 $Mpc/h$. For simplicity, we refer to the standard catalog described in Sect. \ref{results} as PhotoCat and to the results of this run with hybrid redshifts as SpecCat. The comparison between the two catalogs resulted in  1501 successfully matched candidates in total, of which 1404 had $0.03 \leq z \leq 3.7$ and $\lambda_\star > 2$, that is $\sim$ 90\% and $\sim$ 83\% of SpecCat and PhotoCat, respectively. 
Then, we looked for possible unmatched detections that have significantly changed their redshift by introducing spec-$z$s but are still detected above the minimum signal-to-noise. To do so, we matched unmatched detections from SpecCat and PhotoCat, without using the redshift, within a sky separation of 1 arcmin. There are 62 detections in the considered redshift range that are compatible with having the same position in the sky but having a different redshift. We then compared the assigned members to clean from random matches. Among the 62 matches, we found only 4 detections sharing more than 5 associations with $P>20\%$ changing their redshift of a scatter larger than $0.07(1+z)$, so we attribute most of them to random matches.
Then, we analyzed the possibility of redshift fragmentation, namely the possibility that a given detection with a sufficient number of members with spectroscopic redshifts and a sufficient number of members with only photometric redshifts ends up being split into the spectroscopic and photometric components, resulting in two distinct detections aligned along the line of sight. This can be studied by looking at the possible correspondence on the sky plane (without redshift information) between a couple of matched detections and an unmatched spectroscopic detection since this would indicate that the photometric component is still successfully identified but it also gave origin to a new detection at a different redshift due to the spectroscopic galaxies. Among the list of 151 sky matches within 1 arcmin, we found 42 detections that share at least three members with $P>50\%$. However, the selected pairings have a redshift scatter between the matched couple and the only-spectroscopic detection smaller than the average photo-$z$ error. Therefore, in our sample, the redshift fragmentation does not affect the group detection. Finding several examples of pairs coupling with a third detection in the SpecCat may instead indicate that introducing spectroscopic redshifts possibly contributes to reducing over-merging, and the algorithm can more easily distinguish between close-by objects.

\end{appendix}

\end{document}